\documentclass[a4paper,10pt]{article}
\pdfoutput=1 
\usepackage{ae,aecompl}
\usepackage{jheppub} 

\usepackage[T1]{fontenc}
\usepackage[none]{hyphenat}

\usepackage[italian,english]{babel}
\usepackage{hyperref}
\usepackage{changepage}
\usepackage{ifpdf}
\usepackage{subfigure}
\usepackage{amssymb}
\usepackage{amsfonts}
\usepackage{epsf}
\usepackage{rotating}
\usepackage{graphicx}
\usepackage{amsmath}
\usepackage{fancyhdr}
\usepackage{lineno}
\usepackage{babel}
\usepackage{graphics}
\usepackage{pstricks}
\usepackage{color}
\usepackage{multirow}
\usepackage{slashed}
\usepackage{physics}
\usepackage{caption}
\usepackage{tabularx, array,booktabs, makecell}
\usepackage[toc,page]{appendix}
\usepackage{pifont}
\usepackage{orcidlink}    
\usepackage{bm}

\usepackage{tikz}
\usepackage{tikz-feynman}
\tikzfeynmanset{compat=1.0.0}

\definecolor{pantoneCB}{rgb}{0.0588235, 0.298039, 0.505882}

\hypersetup{bookmarks=true,unicode=true,pdftoolbar=true,pdfmenubar=true,
	pdffitwindow=false,
	pdfnewwindow=true,colorlinks=true,allcolors=pantoneCB}

\newcommand{\nn}{\nonumber}
\newcommand{\lsim}{\mathrel{\mathop{\kern 0pt \rlap
			{\raise.2ex\hbox{$<$}}}
		\lower.9ex\hbox{\kern-.190em $\sim$}}}
\newcommand{\gsim}{\mathrel{\mathop{\kern 0pt \rlap
			{\raise.2ex\hbox{$>$}}}
		\lower.9ex\hbox{\kern-.190em $\sim$}}}

\newcommand{\be}{\begin{equation}}
	\newcommand{\ee}{\end{equation}}
\newcommand{\bea}{\begin{eqnarray}}
	\newcommand{\eea}{\end{eqnarray}}

\def\hpm{H^\pm}
\def\tb{\tan\beta}
\def\sba{\sin(\beta-\alpha)}

\def\z2{\mathbb{Z}_2}

\newcommand{\la}{\lambda }

\newcommand{\fbi}{fb$^{-1}$}

\newcommand{\wt}{\widetilde}

\newcolumntype{P}[1]{>{\centering\arraybackslash}p{#1}}
\newcolumntype{M}[1]{>{\centering\arraybackslash}m{#1}}

\title{Probing Boosted Light Scalars in the Type-I 2HDM}
\author[a]{Partha Konar\,\orcidlink{0000-0001-8796-1688},}
\author[b]{Tanmoy Mondal\,\orcidlink{0000-0002-6441-7687},}
\author[a]{and Chandrima Sen\,\orcidlink{0000-0003-4520-1263}}

\affiliation[a]{Theoretical Physics Division, Physical Research Laboratory, Ahmedabad, 380009, Gujarat, India}
\affiliation[b]{Birla Institute of Technology and Science, Pilani, 333031, Rajasthan, India}
\emailAdd{konar@prl.res.in}
\emailAdd{tanmoy.mondal@pilani.bits-pilani.ac.in}
\emailAdd{chandrima@prl.res.in}

\abstract{
In the Type-I two-Higgs Doublet Model (2HDM), the additional scalars may be light ($\lesssim 100$ GeV) without conflicting with experimental constraints from LHC searches or from flavour observables. So far, the studies of light scalars at the LHC have been limited to exploring non-standard decays of the Standard Model (SM) Higgs boson or via $b\bar b$ associated production followed by leptonic decays of the light scalar. A light scalar in Type-I 2HDM can evade these search strategies due to its potentially tiny coupling to the SM Higgs boson and its suppressed coupling to quarks. In this work, we study the electroweak production of scalar pairs involving the light scalar $h$, the pseudoscalar $A$ and the charged Higgs boson $H^\pm$. The subsequent decays of the heavy scalars into $h$ and an electroweak gauge boson lead to multi-$h$ final states, in which the light scalars are typically boosted owing to the hierarchical scalar spectrum. The decay of the boosted $h$ into $b\bar b$ can be reconstructed within a fat-jet containing a pair of $b$-subjets.  We find that tagging such a `boosted double-$b$ fat-jet ($J_{bb}$)' signature in association with a SM gauge boson provides an excellent probe of the Type-I 2HDM for hierarchical scalar spectra. Using multiple light mass $M_h$ benchmarks, we demonstrate that such analysis can explore a large region of the parameter space, with the $2\sigma$ exclusion reach for the heavy scalars extending up to $\sim 560$ GeV ($\sim$ 370 GeV) at the HL-LHC with 3000~\fbi (LHC with 300~\fbi) luminosity for light scalar masses in the range $30$--$70$ GeV. Furthermore, we show that significant sensitivity and even resonance reconstruction can be achieved within a model-independent framework, highlighting the robustness of this search strategy.
}

\allowdisplaybreaks

\begin{document}
\maketitle

\section{Introduction}
The discovery of the Higgs boson~\cite{ATLAS:2012yve, CMS:2012qbp} at the Large Hadron Collider (LHC) put the Standard Model (SM) 
of particle physics on a firm footing. Current LHC data show that the properties of the discovered Higgs boson are consistent with those in the SM~\cite{ATLAS:2022vkf,CMS:2022dwd}. 
However, there are several reasons, both theoretical (the hierarchy problem, gauge coupling unification) and experimental (neutrino masses, the matter-antimatter asymmetry, dark matter, etc.), to believe that the SM is a low-energy effective theory of some more fundamental yet to be discovered theory. 

Light scalar or pseudoscalar states appear in various theoretically well-motivated extensions of the Standard Model. Examples include the Next-to-Minimal Supersymmetric Standard Model (NMSSM)~\cite{Dedes:2000jp,Dobrescu:2000yn,Ellwanger:2003jt,Dermisek:2005ar}, dark-matter portal scenarios~\cite{Pospelov:2007mp,Draper:2010ew,Ipek:2014gua,Arhrib:2013ela}, and models with extended Higgs sectors such as multi-Higgs doublet models~\cite{Lee:1973iz,Branco:2011iw}. Such light scalars can affect the phenomenology of the SM-like Higgs boson, in particular if the SM-like Higgs boson is allowed to decay into them. In this case, the total Higgs width and the measured signal strengths are modified. 
At the LHC, the CMS and ATLAS collaborations have primarily focused on two strategies for the search for 
a light scalar in the mass range of 20 to 70 GeV.  
 The first strategy involves searching for non-standard decays of the 125 GeV Higgs
boson to a pair of light states $AA$\footnote{These discussions are not specific to a pseudoscalar $(A)$ and apply more generally to light bosonic states with suppressed fermionic couplings.}, where the light scalars further decay to SM fermions, 
$b\bar b, \tau^+\tau^-$ or $\mu^+ \mu^-$~\cite{CMS:2018qvj,CMS:2019spf,CMS:2021pcy,CMS:2024uru,CMS:2024zfv,ATLAS:2018jnf,ATLAS:2018pvw,ATLAS:2018emt,ATLAS:2020ahi,ATLAS:2021hbr,ATLAS:2021ldb,ATLAS:2024vpj}. In some cases, the light state might be fermiophobic~\cite{Arhrib:2017uon,Arhrib:2017wmo, Wang:2021pxc, Kim:2023lxc}, which opens up the decay of $A \to \gamma\gamma$, and such final states have also been explored at the LHC~\cite{CMS:2022xxa,CMS:2023eos,ATLAS:2023ian}. The second strategy relies on the production of light scalars in association with $b\bar b$, followed by the $\tau^+\tau^-$ decay of $A$~\cite{CMS:2019hvr}.
In addition, the ATLAS collaboration has also studied gluon fusion production 
of a light pseudoscalar decaying into a pair of tau leptons~\cite{ATLAS:2024rzd}. 
Current experimental data constrain the branching ratio of non-SM Higgs decays to be below about 12\% from ATLAS~ \cite{ATLAS:2022vkf} and 16\% from CMS~\cite{CMS:2022dwd}. With increased luminosity and improved precision, indirect constraints from Higgs coupling measurements at the LHC are expected to probe non-SM branching ratios at the level of a few percent~\cite{Liss:2013hbb,CMS:2013xfa}. However, in many realistic scenarios, the branching ratio of the SM-like Higgs boson into light scalars can be strongly suppressed, well below the percent level, making searches based on Higgs decays ineffective. This motivates the exploration of alternative production mechanisms and search strategies for light scalar states.

In this work, we focus on the Two-Higgs-Doublet Model (2HDM)~\cite{ Gunion:1989we,Davidson:2005cw,Djouadi:2005gj,Branco:2011iw}, and in particular the Type-I 2HDM, as a representative framework in which light scalar states can naturally arise while remaining consistent with current experimental constraints. The 2HDM is one of the simplest extensions 
of the scalar sector, where two Higgs doublets take part in electroweak (EW) symmetry breaking. The scalar spectrum of a $CP$ conserving 2HDM scenario consists of two neutral $CP$-even scalars
(denoted as $h$ and $H$), one $CP$-odd pseudoscalar $A$ and a charged Higgs $\hpm$. We will always assume that $h\,(H)$ is the lighter (heavier) one with mass $M_h (M_H)$. The most general 2HDM suffers from the flavour-changing neutral Higgs interactions, 
and one of the ways to avoid such interactions is to introduce a 
discrete  ($\mathbb{Z}_2$) symmetry \cite{Glashow:1976nt, Paschos:1976ay}, which essentially 
restricts Yukawa interactions. Based on different Yukawa interactions, four distinct 
2HDM, viz., type I, II, X, and Y, are possible, which give rise to different phenomenologies. The Type-I 2HDM is defined  such that fermion fields transform oddly under the $\mathbb{Z}_2$
symmetry and therefore couple only to one of the Higgs doublets. Due to such a Yukawa structure, the Beyond the SM (BSM) scalars couple to fermions with a suppression proportional to $\tb$, the ratio of 
vacuum expectation values of the two Higgs doublets. Consequently, their contributions to flavour-changing and charged-current processes, such as $b\to s\gamma$, meson mixing, and leptonic meson decays, are 
significantly suppressed. Hence, the BSM scalars in the Type-I 2HDM remains weakly constrained 
and can be quite light, and the neutral one can even lie below $62.5$ GeV, without conflicting with current flavour physics bounds. 
Within the Type-I 2HDM, the presence of a light scalar state can be realized in two ways. One possibility corresponds to a light $CP$-odd scalar $A$, while the remaining non-SM Higgs states are comparatively heavy. The other realization is the so-called `inverted hierarchy', where the heavier $CP$-even scalar ($H$) is identified with the observed 125 GeV SM-like Higgs boson, while the lighter $CP$-even state $h$ lies below the electroweak scale.

Despite extensive LHC searches for SM Higgs decay to light scalars into various final states, these analyses cannot exclude the Type-I 2HDM with light $h$ or $A$. This happens because in the Type-I 2HDM, the coupling of the light state with the SM-like Higgs boson depends on a combination of parameters of the scalar potential, which can be arbitrarily small. Moreover, the gluon fusion-initiated processes can be subdominant as the fermionic coupling of the light scalar is $\tb$ suppressed \cite{Akeroyd:1995hg, Akeroyd:1998ui, Arhrib:2016wpw, Enberg:2018pye, Kling:2020hmi, Wang:2021pxc, Bahl:2021str, Arhrib:2021xmc, Mondal:2021bxa, Kim:2022nmm, Kim:2023lxc, Coutinho:2024zyp}. Note that, while much of the existing experimental literature referred here is phrased in terms of a light pseudoscalar $A$, the discussion here is more general and stems from the presence of a light bosonic state with suppressed couplings to fermions. 

From the above discussion, it is evident that we must explore other 
ways to conclusively probe the Type-I 2HDM parameter space which allows such a light (pseudo)scalar. The 
electroweak production of BSM Higgs states, governed by the gauge coupling, is one such possibility and
has the potential to provide a stronger limit than the existing one coming from LHC~\cite{Enberg:2018pye, Arhrib:2021xmc, Mondal:2021bxa}. In the EW production mode, the light (pseudo)scalar is produced in association with a heavy Higgs $(H)A$ or a charged Higgs $\hpm$.
The subsequent decay of the heavy states into the (pseudo)scalar one via the gauge coupling 
gives rise to a multi-$A$ or multi-$h$ signature, often augmented by an SM gauge boson\cite{Arhrib:2021xmc, Kang:2022mdy, Li:2023btx}. Such a state is b-jet rich since the light (pseudo)scalar predominantly decays to $b\bar{b}$. 
Recently, it has been demonstrated that EW production of such multi-$A$ final states can simultaneously yield visible signatures of all three BSM Higgs bosons in the Type-I 2HDM at the LHC~\cite{Mondal:2023wib}. EW production of the charged Higgs with a light $A$ has been studied in ~\cite{Sanyal:2023pfs}. Phenomenological analysis of light $h$ states in Type-I 2HDM has been done in ref~\cite{Arhrib:2023apw}.

In these studies, not only is the (pseudo)scalar light ($\leq 70$ GeV), but the remaining BSM Higgs states are also required to be not too heavy in order to maintain an observable production rate. 
As the mass gap between the light state and the heavier Higgs states ($A$ or $\hpm$) increases, the light state is produced with large transverse momentum ($p_T$). 
This boosted regime is a generic feature of scenarios with a large mass hierarchy. Consequently, the decay products of the light state become collimated and reconstructed as a single fat-jet rather than as two resolved $b$-tagged jets.
Since existing phenomenological analyses rely predominantly on resolved $b$-jet topologies, their sensitivity degrades rapidly as the mass hierarchy increases. We would like to emphasize that the presence of such a boosted regime is a generic consequence of a hierarchical scalar spectrum and current phenomenological studies are unable to probe regions of the Type-I 2HDM parameter space with a significant mass separation, despite these regions being theoretically viable. 

Given that the recent machine-learning-based flavour tagging algorithms can tag multiple $b$-jets within a fat-jet with reasonable accuracy~\cite{Moreno:2019neq,Lin:2018cin,Ghosh:2026rcf}, it is worthwhile to explore how such techniques can be useful to probe yet unexplored regions in 2HDM. The combination of electroweak production and ML-based boosted-object tagging, therefore, provides a unique opportunity to close one of the remaining gaps in the collider exploration of the Type-I 2HDM. Motivated by this, we propose a complementary search strategy targeting scenarios with a light scalar and a large mass gap among the BSM scalars. 

We demonstrate that a pair of large radius jets, each containing a pair of $b$-subjets, will be an excellent probe for a scalar spectrum with a large mass gap. Based on the decay topology and the number of fat-jets, we identify four final states relevant for probing the inverted Type-I 2HDM. We show that it is possible to minimize the SM background processes significantly, and thus can probe the Type-I 2HDM to a much higher mass scale compared to existing approaches based on resolved $b$-jets. We perform a detailed analysis for several benchmark scenarios to demonstrate  the robustness of our signal in the boosted regime. Apart from the benchmark searches, we have also explored a model agnostic search strategy that is independent of the specific masses of the BSM scalars. 
Although benchmark analysis provides a higher significance, the model agnostic search strategy remains robust  and provides significant bound on the 2HDM parameter space. Together, these analyses demonstrate that the double-b fat-jet offers a complementary probe of the inverted Type-I 2HDM scenario. We would like to emphasize that the sensitivity of the proposed signal has a weak dependence on $\tb$, allowing it to probe a large region of the parameter space, particularly at high $\tb$ and in mass ranges that remain inaccessible to existing search strategies.

The article is organized as follows: in \autoref{sec:2-model}, we briefly review the Type-I 2HDM. The parameter space relevant to our study, together with the applied theoretical and experimental constraints, is discussed in \autoref{sec:3-bp}. The collider phenomenology of the boosted light scalar, along with the details of the event generation, analysis framework for four possible final states, and relevant backgrounds, is presented in \autoref{sec:4-pheno}. The cut-based analysis results for the most relevant final state and sensitivity projections for all four final states are given in \autoref{sec:5-result} and \autoref{sec:model_indep}, corresponding to the model-dependent and model-independent analyses, respectively. We summarize and conclude our findings in \autoref{sec:6-concl}. Additional information for other final states is provided in \autoref{app:cutflow}.

\section{The Type-I 2HDM Model}\label{sec:2-model}
In the 2HDM~\cite{Gunion:1989we,Davidson:2005cw,Djouadi:2005gj,Branco:2011iw}, the SM scalar sector is augmented by another scalar doublet with hypercharge +1/2.
Due to the presence of two Higgs doublets, the fermions can couple to both 
the scalar doublets, giving rise to flavour-changing neutral current (FCNC) 
processes which are highly constrained by experiments. To avoid such processes, one needs to restrict the coupling of the fermions to only one of the doublets. We ensure this by introducing a $\z2$ symmetry~\cite{Glashow:1976nt, Paschos:1976ay} under which the scalars are oppositely charged, $\Phi_1 \to \Phi_1$ and
$\Phi_2\to -\Phi_2$, and the SM fermions are charged appropriately.
\subsection{Scalar Sector}
The general form of the scalar potential obeying\footnote{In the scalar potential, we allow terms like $m_{12}^2$ which breaks the $\z2$ symmetry softly. It also keeps $\la_1$ under the perturbative limit.} $\z2$ symmetry is given by,
\begin{eqnarray}
\nonumber V_{\mathrm{2HDM}} &=& -m_{11}^2\Phi_1^{\dagger}\Phi_1 - m_{22}^2\Phi_2^{\dagger}\Phi_2 -\Big[m_{12}^2\Phi_1^{\dagger}\Phi_2 + \mathrm{h.c.}\Big]
+\frac{1}{2}\lambda_1\left(\Phi_1^\dagger\Phi_1\right)^2+\frac{1}{2}\lambda_2\left(\Phi_2^\dagger\Phi_2\right)^2 \\
 && +\lambda_3\left(\Phi_1^\dagger\Phi_1\right)\left(\Phi_2^\dagger\Phi_2\right)+\lambda_4\left(\Phi_1^\dagger\Phi_2\right)\left(\Phi_2^\dagger\Phi_1\right)
+\Big\{ \frac{1}{2}\lambda_5\left(\Phi_1^\dagger\Phi_2\right)^2+ \rm{h.c.}\Big\}.
\label{eq:2hdm-pot}
\end{eqnarray}
We also assume a $CP$-conserving scenario which ensures that $m_{12}^2$ and $\la_5$
are real parameters, and the $\mathbb{Z}_2$ symmetry is softly broken by the dimensionful coupling $m_{12}^2$. The Higgs doublets are parameterized in the following way,
\be 
\Phi_j=\begin{pmatrix}
        H_j^+\\\dfrac{1}{\sqrt2}(v_j + h_j + i A_j) 
       \end{pmatrix},\hspace{1cm} j = 1,2.
\ee
and define $\tb=\frac{v_2}{v_1}$. After spontaneous symmetry breaking, the scalar 
sector consists of two $CP$-even Higgs bosons, $h$ and $H$, one $CP$-odd pseudoscalar $A$ and a charged Higgs boson $\hpm$.
The mass eigenstates are related to gauge eigenstates via the following rotations,
\begin{align}\label{2hdm_scalar_basis}
 \begin{pmatrix} H  \\ h \end{pmatrix} =  \begin{pmatrix}  c_{\alpha} && s_{\alpha} \\ -s_{\alpha} &&  c_{\alpha} \end{pmatrix}
  \begin{pmatrix} h_1  \\ h_2 \end{pmatrix}, \quad\textrm{and}~~ A(\hpm)=-s_\beta \;A_1(H_1^{\pm}) + c_\beta \;A_2 (H^{\pm}_2), 
 \end{align}
where we denoted $s_\alpha = {\rm sin}~\alpha$, $c_\beta = {\rm cos}~ \beta$ etc.
The scalar potential contains a total eight free parameters, i.e three quadratic 
mass-squared terms and 5 quartic couplings. We can trade them for the following set 
of phenomenologically relevant parameters,
\be
    M_h^2,\,M_H^2,\,M_A^2,\,M_{\hpm}^2,\,\tb,\,\sin(\beta-\alpha),\,m_{12}^2 \textrm{ and } v.
 \ee
 Moreover, we can also treat $\la_1$ as a free parameter instead of $m_{12}^2$, related via the following relation,
 \be
m_{12}^2 =
\frac{
M_H^2 \cos^2\alpha
+
M_h^2 \sin^2\alpha
-
\lambda_1 v^2 \cos^2\beta
}{\tan\beta}.
 \ee
In this work, we are interested in scenarios where one scalar is lighter than the 
SM Higgs and other BSM scalars are heavier. Based on the masses of the BSM scalars, two main possibilities emerge in Type-I 2HDM. 

 \subsubsection*{Light Pseudoscalar Scenario}
In this scenario, $h$ is identified as the SM-like Higgs boson with mass 125 GeV, whereas $H$ is the heavier one. Also, the pseudoscalar is assumed to be lighter than 125 GeV. 
For such a scalar spectrum, the triple scalar couplings $\la_{hAA}$ dictates the exotic decay of the SM Higgs boson. The coupling is given by,
\begin{eqnarray}
\lambda_{hAA} 
&=&
\frac{1}{4v^2 s_\beta c_\beta}
\left[
\left(4 \frac{m_{12}^2}{\sin\beta\cos\beta} - 2M_A^2 - 3M_h^2\right)\cos(\alpha+\beta)
+ (2M_A^2 - M_h^2)\cos(\alpha-3\beta)
\right]
\end{eqnarray}
It is possible to tune the free parameter $m_{12}^2$ to achieve a very small exotic branching ratio 
of the SM Higgs boson. However, once we fix $m_{12}^2$, the quartic coupling $\lambda_1$ 
becomes a dependent parameter which grows as $M_H$ increases. We found that for $\tb = 3$, the condition that $\la_1 \leq 4\pi$ requires $M_H < 270$ GeV and the limit grows stronger as we increase $\tb$. 
Thus, in this light $A$ scenario, it is not possible to achieve a large mass hierarchy in Type-I 2HDM. 
Hence, we have not considered this scenario for our analysis.

\subsubsection*{Light Scalar Scenario: Inverted Type-I 2HDM}
\begin{center}
    \begin{figure}
        \centering
        \includegraphics[width=0.7\linewidth]{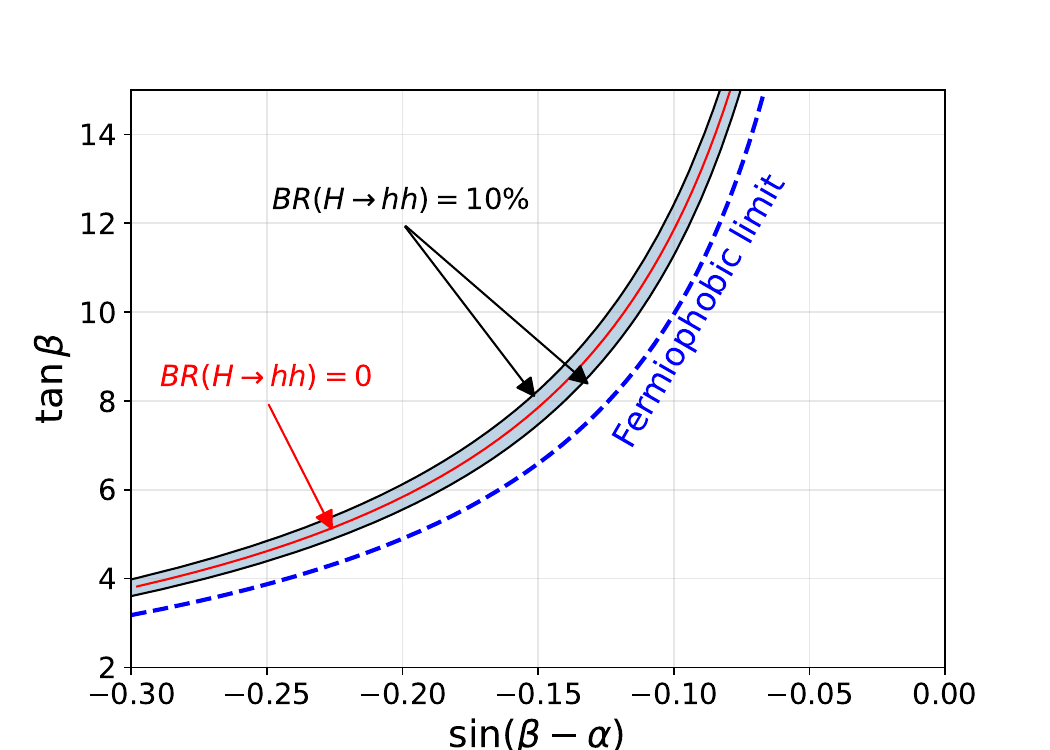}
        \caption{The shaded region shown where the branching ratio of the SM Higgs to light scalars is less than 10\% whereas the central red curve shows $\Gamma_{Hhh}$ = 0. The dashed blue curve depicts the `fermio-phobic' limit where the light Higgs $h$ does not couple to fermions and is excluded by SM Higgs data.  In this plot we fix $M_h=50$ GeV.}
        \label{fig:br-Hhh}
    \end{figure}
\end{center}

In the \emph{inverted hierarchy} scenario, the heavier of the two $CP$-even scalars is identified as the SM Higgs boson ($H$) with mass 125 GeV, while the other $CP$-even scalar $h$ is lighter. The pseudoscalar ($A$) and the charged Higgs ($H^\pm$) are heavier than the SM Higgs. Henceforth, we will use this notation for the inverted scenario throughout the text, unless specified otherwise.

The SM Higgs can decay to the 
light scalar $h$, and the corresponding coupling is given by,
\begin{equation}\label{eq:Hhh}
\lambda_{Hhh}=\frac{1}{v}\left(\frac{\cos\alpha}{\sin\beta}+\frac{\sin\alpha}{\cos\beta}\right)
\left[-\,m_{12}^2-\cos\alpha\,\sin\alpha
\left(M_H^2+ 2 M_h^2-\frac{3\,m_{12}^2}{\sin\beta\,\cos\beta}
\right)\right] .
\end{equation}
As before, we can leverage $m_{12}^2$ to keep the exotic branching ratio of the SM Higgs boson below the permissible limit. In \autoref{fig:br-Hhh} we have shown the parameter space in the $\sba-\tb$ plane where $BR(H\to hh) < 10\%$. The red curve in the center of the colored region depicts where the coupling $\lambda_{Hhh}$ in \autoref{eq:Hhh} vanishes. It is worth noting that only negative values of $\sba$ are allowed by the SM Higgs data. In this plot, we chose $M_h = 50$ GeV and $m_{12}^2 = M_h^2 \sin\beta\,\cos\beta$. In addition, we also plot the 
`fermio-phobic' limit where the coupling of the lighter scalar to fermions ($\xi_h^f$) as shown in \autoref{Tab:YukawaFactors} vanishes. Such a limit is excluded by the current SM Higgs data as the parameter space with fermio-phobic limit will violate exotic decay branching ratio of the SM Higgs boson. 
Away from this limit, all fermionic decay widths are governed by the common Type-I Yukawa factor \(\cos\alpha/\sin\beta\). For \(M_h=50-70~{\rm GeV}\), we typically get
\[
{\rm BR}(h\to b\bar b)\simeq86\%,
\qquad
{\rm BR}(h\to\tau^+\tau^-)\simeq8\%.
\]
Evidently, lower branching fraction and the missing energy carried by neutrinos reduce the signal yield and mass-reconstruction efficiency in \(\tau^+\tau^-\) mode. Similarly, decay into diphoton mode can become relevant near the fermiophobic limit, but remains strongly suppressed in the allowed parameter region considered here.
Hence, within the allowed parameter space (the shaded region in \autoref{fig:br-Hhh}), the light scalar $h$ predominantly decays to $b\bar{b}$.

Moreover, in contrast to the light $A$ scenario, the quartic coupling $\la_1$ remains small as it is associated with $h$, the lightest scalar.
Thus, in this scenario, it is possible to have a large mass hierarchy among the BSM scalars, and we will be considering this scenario for our analysis. 

\subsection{Yukawa Sector}
\begin{table}[t]
\begin{center}
\begin{tabular}{|c||c|c|c||c|c|c||c|c|c|}
\hline
2HDM& $\xi_h^u$ & $\xi_h^d$ & $\xi_h^\ell$
& $\xi_H^u$ & $\xi_H^d$ & $\xi_H^\ell$
& $\xi_A^u$ & $\xi_A^d$ & $\xi_A^\ell$ \\ \cline{2-10}

Type-I& $c_\alpha/s_\beta$ & $c_\alpha/s_\beta$ & $c_\alpha/s_\beta$
& $s_\alpha/s_\beta$ & $s_\alpha/s_\beta$ & $s_\alpha/s_\beta$
& $\cot\beta$ & $-\cot\beta$ & $-\cot\beta$ \\
 \hline
\end{tabular}
\end{center}
 \caption{The Yukawa multiplicative factors in Type-I 2HDM}
\label{Tab:YukawaFactors}
\end{table}

As we have mentioned earlier, the $\z2$ symmetry restricts the coupling of the 
fermions with the scalar doublets, ensuring that tree-level FCNC interactions vanish.
Based on the Yukawa interactions, for distinct possibilities arise and in Type-I
2HDM, all the right-handed particles of the SM are odd under the $\mathbb Z_2$ symmetry and thus only couple 
to $\Phi_2$. The Yukawa Lagrangian is given by, 
\be\label{eq:yukawa}
-{\cal L}_Y= Y^u\bar{ Q_L} \wt \Phi_2 u_R + Y^d  \bar{ Q_L} \Phi_2 d_R+Y^e\bar{ l_L} \Phi_2 e_R + h.c.. 
\ee
After EWSB, we can write the interaction of the fermions with the scalars in the mass basis,
\begin{eqnarray}
\nonumber \mathcal L_Y^{\mathrm{mass}} &=&
-\sum_{f=u,d,\ell} \frac{m_f}{v}\left(\xi_h^f\overline{f}hf +
\xi_H^f\overline{f}Hf - i\xi_A^f\overline{f}\gamma_5Af \right) \\
 &&-\left\{ \frac{\sqrt{2}V_{ud}}
{v}\overline{u}\left(\xi_A^{u} m_{u} P_L+\xi_A^{d} m_{d} P_R\right)H^{+}d  +
\frac{\sqrt{2}m_l}{v}\xi_A^l\overline{v}_LH^{+}l_R + \mathrm{h.c.}\right\}.
\label{eq:L2hdm}
\end{eqnarray}
Here, the quarks and charged leptons are denoted as $u$, $d$, and $l$, and $V$ is the CKM matrix, whereas $P_{L(R)}$ are the chiral projection operators.
The Yukawa multiplicative factors $\xi^f_\phi$ in \autoref{eq:L2hdm} depend on the exact nature of Yukawa 
couplings, and for the Type-I scenario, the corresponding coupling modifiers are shown in \autoref{Tab:YukawaFactors}. It is easy to see that all the couplings of the 
BSM scalars with the fermions are inversely proportional to $\tb$ and thus 
new scalar couples weakly to fermions for large $\tb$.

\subsection{Gauge Sector}
The kinetic term of the Higgs doublets gives the interaction of the scalar particles with 
the gauge bosons. The gauge-Higgs interactions are given by,
\be \label{eq:L-gauge}
\mathcal L_{\rm int} =
\sum_a \kappa_V^a g_{HVV}^{\rm SM} H_a V_\mu V^\mu +
\sum_{V,a<b} g_{V H_a H_b}
V_\mu H_a \!\stackrel{\leftrightarrow}{\partial^\mu}\! H_b,
\ee
where $H_a=\{h,H,A,H^\pm\}$ denote the physical Higgs states and 
$V=\{Z,W^\pm\}$ denotes the electroweak gauge bosons. The first term in 
\autoref{eq:L-gauge} describes the couplings of the $CP$-even scalars to 
with gauge boson pairs, normalized to the SM coupling  
$g_{HVV}^{\rm SM}=2m_V^2/v$, with scaling factor
$\kappa_V^h=\sin(\beta-\alpha)$ and $\kappa_V^H=\cos(\beta-\alpha)$,
while $\kappa_V^A=\kappa_V^{H^\pm}=0$. The second term denotes the
gauge--Higgs--Higgs interactions, and in the $CP$-conserving limit, the only non-vanishing couplings are
$Z h A$ and $Z H A$, proportional to $\cos(\beta-\alpha)$ and
$-\sin(\beta-\alpha)$, respectively. Moreover, the charged-current
interactions include $W^\pm H^\mp h$, $W^\pm H^\mp H$, and
$W^\pm H^\mp A$, where the first two are proportional to
$\cos(\beta-\alpha)$ and $-\sin(\beta-\alpha)$, respectively, and the
$W^\pm H^\mp A$ coupling is independent of the $CP$-even mixing angle. 
All other gauge--Higgs--Higgs couplings vanish at tree level by $CP$- 
and gauge invariance.

\section{Constraints}\label{sec:3-bp}
Any viable parameter space of the Type-I 2HDM must satisfy the theoretical and experimental constraints. In this section, we briefly summarise the most relevant constraints.

\subsection*{Theoretical Constraints}

\begin{itemize}
    \item We impose the condition that all the quartic couplings must
    remain perturbative, $|\lambda_i| < 4\pi$.
    \item The scalar potential of the 2HDM must be bounded from below to ensure the existence of a stable vacuum. This requirement imposes positivity conditions on the quartic couplings of the scalar potential given by~\cite{Gunion:2002zf},
\be
\lambda_1 > 0, \qquad \lambda_2 > 0, \qquad \lambda_3 > -\sqrt{\lambda_1 \lambda_2}, 
\qquad\lambda_3 + \lambda_4 - |\lambda_5| > -\sqrt{\lambda_1 \lambda_2}.
\ee
\item Perturbative unitarity of scalar--scalar $2\to 2$ scattering amplitudes places upper bounds on combinations of the quartic couplings.  In the $\mathbb{Z}_2$ symmetric 2HDM, the analytic expressions for the eigenvalues of the $s$-wave amplitude matrix are discussed in Refs.~\cite{Kanemura:1993hm,Akeroyd:2000wc,Horejsi:2005da}. We require the eigenvalues of the $s$-wave scattering matrix to satisfy $|a_0| < \frac{1}{2}$.

\end{itemize}

\subsection*{Electroweak Precision Constraints }
Electroweak precision constraints are imposed using the oblique parameters
$S$, $T$, and $U$~\cite{Peskin:1991sw}.  In our analysis, we set $U=0$, which is a well-motivated
approximation for models in which custodial symmetry is only mildly broken.
The predicted values of $S$ and $T$ in the 2HDM are computed using
\texttt{2HDMC-1.8.0}~\cite{Eriksson:2009ws}, and compared against the experimentally preferred values
$S_0 = 0.02, ~ T_0 = 0.06$, 
with corresponding one-standard-deviation uncertainties
$\sigma_S = 0.07, ~ \sigma_T = 0.06$,
and correlation coefficient $\rho = 0.92$, as obtained from global electroweak
fits~\cite{ParticleDataGroup:2022pth}. The compatibility with electroweak precision data is quantified through a
$\chi^2$, 
\be
\chi^2_{ST}
=
\begin{pmatrix}
S-S_0 & T-T_0
\end{pmatrix}
\mathbf{C}^{-1}
\begin{pmatrix}
S-S_0 \\
T-T_0
\end{pmatrix},\quad
\textrm{where}\quad 
\mathbf{C} =
\begin{pmatrix}
\sigma_S^2 & \rho\,\sigma_S\sigma_T \\
\rho\,\sigma_S\sigma_T & \sigma_T^2
\end{pmatrix}.
\ee
Model points are retained if $\chi^2_{ST} < 5.99$, corresponding to the
$95\%$ CL. 

\subsection*{Flavor Constraints}
In Type-I 2HDM, the coupling of the charged Higgs to fermions is suppressed by $\tb$, and 
thus is much weaker than the  Type-II scenarios. Processes such as
$B \to X_s \gamma$ place a limit of $\tb > 2$~\cite{HFLAV:2016hnz,Misiak:2017bgg} for any values of the charged Higgs mass.  
\begin{center}
    \begin{figure}
        \centering
        \includegraphics[width=0.7\linewidth]{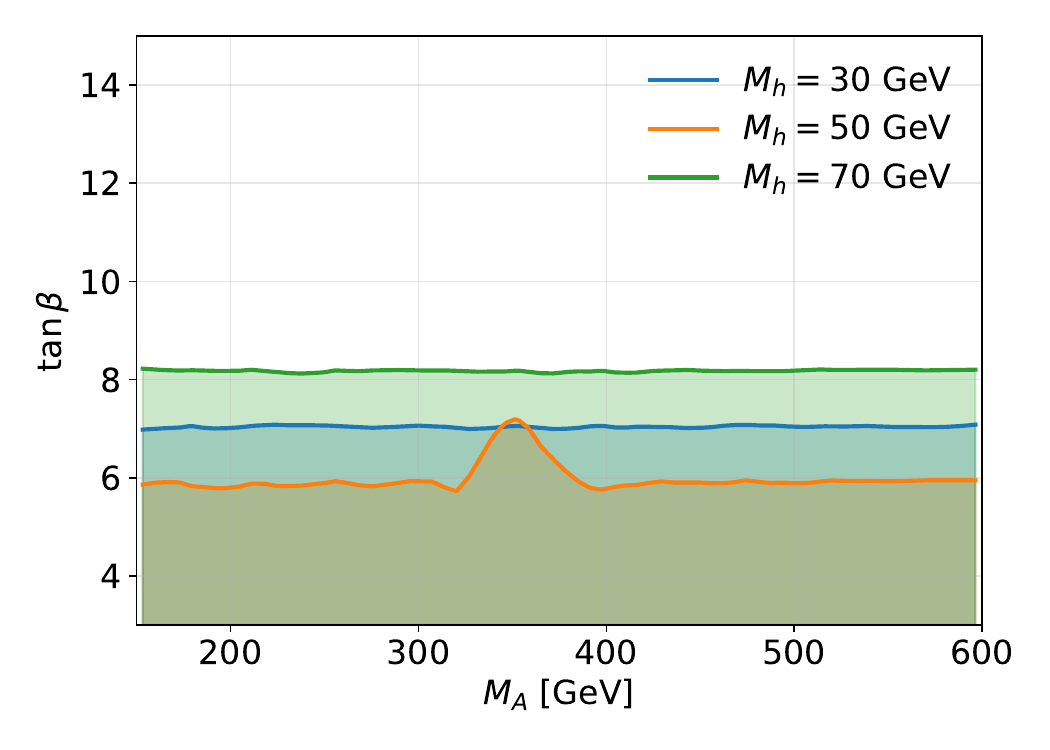}
        \caption{The shaded region is excluded from the direct search of BSM scalars at the 
        LHC and LEP. The three different colours correspond to three different light scalar masses.}
        \label{fig:higgsbound}
    \end{figure}
\end{center}
\subsection*{Experimental Constraints}
Direct searches for additional Higgs bosons and precision measurements of the $125\,\mathrm{GeV}$ Higgs boson properties at the LHC impose strong constraints on the 2HDM-I parameter space. 
For the Higgs signal strength, the dominating constraint comes from the exotic decay of the 
Higgs boson, which can be suppressed as shown in  \autoref{fig:br-Hhh}.

To estimate bounds coming from direct searches at the LHC, we use {\tt HiggsBounds-v5.10.2} \cite{Bechtle:2020pkv}. To satisfy the Higgs decay constraints, we fix $m_{12}^2 = M_h^2\sin\beta\cos\beta$ and estimate $\sba$ such that $\lambda_{Hhh} \approx 0$, i.e.,
\be
\sin(\beta-\alpha) = \sin\!\left[ \frac{1}{2} \tan^{-1}\!\left(
\frac{M_H^2}{M_H^2 - M_h^2} \tan(2\beta) \right)\right].
\ee
\autoref{fig:higgsbound} depicts the excluded parameter space in the $M_A$–$\tb$ plane. The shaded regions are excluded from direct searches for BSM scalars at the LHC and LEP. The three different colors correspond to $M_h = 30$, $50$, and $70~\text{GeV}$.

For $M_h = 30~\text{GeV}$ and $70~\text{GeV}$, the strongest limits arise from the LEP searches~\cite{ALEPH:2006tnd,LEP:2003ing}
\be
e^+ e^- \to h\,Z \to b\,\bar b\, Z,
\ee
and thus the limits are independent of the heavy scalar masses.
The CMS analysis~\cite{CMS:2019ogx}
\be
pp \to H/A \to h Z \to (b\bar b)\,(\ell\ell)
\ee
also becomes relevant near the $t\bar t$ threshold, since the production of the heavy pseudoscalar via gluon fusion is enhanced, leading to stronger limits on $\tb$ from the CMS study~\cite{CMS:2019ogx}. This effect is more pronounced for $M_h = 50~\text{GeV}$, where the corresponding LEP limits are relatively weak.

From \autoref{fig:higgsbound}, it is evident that for large $\tb$, the parameter space remains largely unconstrained by current theoretical and experimental bounds. In this work, we propose boosted double-$b$ signatures as an alternative approach to probe this allowed region.

\section{Phenomenology of a Boosted Scalar}\label{sec:4-pheno}

From the above discussion of various constraints, it is evident that in the inverted hierarchy scenario of the Type-I 2HDM, a large region of parameter space with $\tan\beta \gsim 8$ remains viable independent of the heavy BSM scalar masses. Since the coupling of the BSM scalars with fermions in the Type-I 2HDM is $\tan\beta$ suppressed, the dominant production mode of the heavy scalars is via electroweak gauge bosons, and the subsequent decay branching ratios of the heavy scalars ($A, H^\pm$) to the light one $h$ in association with the gauge boson is close to unity. Thus, a hierarchical mass spectrum always produces multiple $h$ in the final state. The light scalar $h$ can further decay to $b\bar b$ or, in the case of the fermiophobic scenario, to $\gamma\gamma$. However, as we discussed above, SM Higgs measurements strongly constrain $\sin(\beta-\alpha)$ and $\tan\beta$ disfavoring the fermiophobic limit. Hence, $h$ always decays predominantly to $b\bar b$. When the mass gap between $h$ and $A$ (or $H^\pm$) is large, the light scalar is produced with a large transverse momentum ($p_T$), which leads to a boosted topology. Consequently, the decay products of $h$ become collimated and can be efficiently reconstructed as a single fat-jet containing two resolved $b$-subjets. 
In this work, we demonstrate that such boosted topologies provide an excellent probe of the inverted Type-I 2HDM scenario, allowing us to probe heavy scalar masses above 500 GeV, far beyond the reach of existing searches, even at the HL-LHC.

To demonstrate the efficacy of our proposed signal of a fat-jet containing two resolved $b$-subjets, we perform a dedicated reconstruction-level analysis for the relevant parameter space in Type-I 2HDM. We vary the light scalar mass in the range $M_{h} \in [30,70]~\mathrm{GeV}$, and the heavier scalar masses are scanned over $M_{A}, M_{H^\pm} \in [150,650]~\mathrm{GeV}$. For simplicity, we fix the mass splitting between the pseudoscalar and the charged Higgs boson to be small, $M_{H^\pm}-M_A=5~\mathrm{GeV}$, and maintain this choice throughout the analysis. We also study the dependence on $\tan\beta$ by scanning the range $\tan\beta=2$--$20$ when presenting the reach in the $M_A$- $\tan\beta$ plane. This setup allows us to cover the phenomenologically relevant hierarchical spectrum while keeping the analysis tractable, without affecting the generality of our conclusions.

In addition to the masses, we also explore the dependence on $\tan\beta$ by varying $\tan\beta$ from 5 to 50. The values of $\sin(\beta-\alpha)$ are chosen such that the SM-like Higgs properties remain consistent with experimental measurements, while the exotic decay $H \to h h$ remains suppressed. The parameter $m_{12}^2$ is fixed accordingly to ensure a viable scalar potential and compatibility with theoretical constraints.

\subsection{Benchmark}
While the collider reach of the signal is obtained from a continuous scan of the parameter space, we select a set of benchmark points (BPs) representative of the viable parameter space to quantify our analysis. The details of the benchmark points are summarized in \autoref{Tab:BPs}. These BPs are chosen to illustrate different mass hierarchies and boost regimes relevant for our analysis.
In particular,  we consider three different values of the light scalar mass, $M_{h} = 30, 50,$ and $70~\mathrm{GeV}$, which lie well below the electroweak scale and are consistent with current experimental constraints. For each choice of $M_{h}$, the masses of the heavier scalars ($A$ and $H^\pm$) are chosen over a wide range to probe different degrees of mass hierarchy, which directly control the boost of the light scalar. In all benchmark points, we fix the mass splitting between the pseudoscalar and the charged Higgs boson to be small, $M_{H^\pm} - M_{A} = 5~\mathrm{GeV}$. We fix $\tan\beta = 10$ for all benchmark points to ensure suppression of the fermionic couplings of the BSM scalars, thereby satisfying flavour constraints. 

The benchmark points introduced here are used to illustrate the reconstruction efficiencies, cut flows, and relative contributions of the different production channels. The final discovery and exclusion sensitivities of the model are not inferred from these discrete points; they are obtained from continuous scans over the relevant parameter space, as presented in the subsequent sections.

\begin{table*}[t]
	\renewcommand{\arraystretch}{1.6}
	\centering
	\hspace*{-0.7cm}
	\scalebox{0.82}
		{\begin{tabular}{ |c|c|c|c|c|c|c||c|c|c|c| }
			\hline				
			\multirow{1.5}{*}{Benchmark} & \multirow{1.5}{*}{$M_{A}$} & \multirow{1.5}{*}{$M_{H^\pm}$} & \multirow{1.5}{*}{$M_{h}$}  & \multirow{1.3}{*}{$m_{12}^2$}  & \multirow{2}{*}{$\sin(\beta - \alpha)$} & \multirow{2}{*}{$\tan \beta$ } & \multicolumn{4}{c|}{Cross-section (in fb)}\\ 
            \cline{8-11}
			\multirow{1}{*}{Points} & \multirow{1}{*}{(GeV)} & \multirow{1}{*}{(GeV)} & \multirow{1}{*}{(GeV)}    & \multirow{1}{*}{(GeV$^2$)} & & & $pp\rightarrow hA$ & $pp\rightarrow hH^\pm$ & $pp\rightarrow AH^\pm$ & $pp\rightarrow H^+H^-$ \\ \hline \hline
			bp30A &  150.0 & 155.0  &  \multirow{2}{*}{30.0} & \multirow{2}{*}{89.0}  & \multirow{2}{*}{-0.107} & \multirow{2}{*}{10.0} & 324.28  & 501.16 & 117.38 & 62.26 \\ 
            \cline{1-3}
            \cline{8-11}
			bp30B & 300.0 & 305.0  & &  &  & & 29.60  & 51.17 & 9.55 & 5.19 \\ 
			\hline \hline
			bp50A & 250.0 & 255.0 & \multirow{2}{*}{50.0}   & \multirow{2}{*}{220.0}  & \multirow{2}{*}{-0.115} & \multirow{2}{*}{10.0} & 50.00 & 85.16 & 18.92 & 10.27 \\ \cline{1-3}
            \cline{8-11}
			bp50B & 350.0 & 355.0 &   &  &  & & 15.91 & 28.07 & 5.26 & 2.85 \\ 
			\hline \hline
			bp70A & 300.0 & 305.0 & \multirow{2}{*}{70.0}  & \multirow{2}{*}{485.0} & \multirow{2}{*}{-0.115} & \multirow{2}{*}{10.0} & 24.28 & 42.38 & 9.55 & 5.19 \\ 
			\cline{1-3}
            \cline{8-11}
			bp70B & 400.0 & 405.0 &  &  &  & & 9.18 & 16.46 & 3.09 & 1.67 \\ 
			\hline
		\end{tabular}}
		\caption{Benchmark points used in the analysis, corresponding to three representative values of the light scalar mass $M_{h} = 30, 50,$ and $70~\mathrm{GeV}$. For each case, the masses of the heavier scalars ($A$ and $H^\pm$) are varied to explore different mass hierarchies. The parameters $\tan\beta$, $\sin(\beta-\alpha)$, and $m_{12}^2$ are chosen to satisfy theoretical constraints, electroweak precision observables, flavor bounds, and Higgs signal strength measurements. Production cross-sections for the signal processes $pp \to h A$, $pp \to h H^\pm$, $pp \to A H^\pm$ and $pp \to H^+ H^-$ at the LHC with $\sqrt{s} = 14$ TeV for the chosen benchmark points are shown in last four columns. The values are computed at leading order and rescaled by a constant $K$-factor of 1.34. } \label{Tab:BPs}
\end{table*}	

\subsection{Signal Characterization}
As discussed above, we consider the electroweak production
of heavy BSM scalars, which leads to multiple boosted light scalar signatures 
via the following processes:
\begin{eqnarray}
	pp&\to& Z^\ast \to h\,A \to h (h\,Z) \to (b\bar b)\, (b\bar b)\, Z, \nn\\
	pp&\to& {W^\pm}^\ast \to h\,H^\pm \to h (h\,W^\pm) \to (b\bar b)\, (b\bar b)\, W^\pm,
    \nn\\
    pp&\to& {W^\pm}^\ast \to A\,H^\pm \to (hZ) (h\,W^\pm) \to (b\bar b)\,Z\, (b\bar b)\, W^\pm,
    \nn\\
    pp&\to& Z^\ast/ \gamma \to H^\pm\,H^\mp \to (h\,W^\pm) (h\,W^\mp) \to (b\bar b)\,W^\pm\, (b\bar b)\, W^\mp.
\end{eqnarray}

\begin{figure}[hbt]
	\begin{center}
		\mbox{\subfigure[]{\includegraphics[width=0.48\linewidth,angle=0]{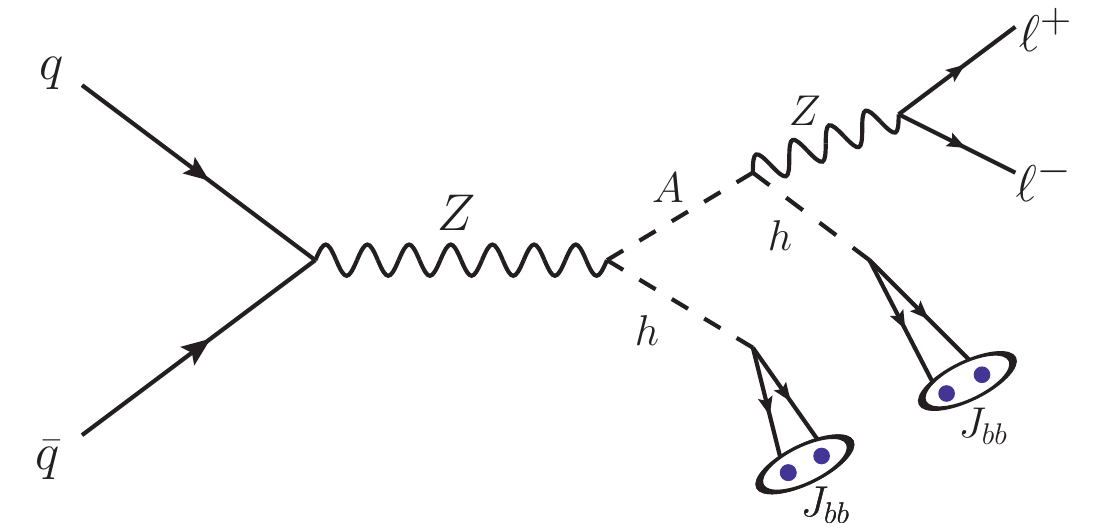}}\quad \quad
			\subfigure[]{\includegraphics[width=0.48\linewidth,angle=0]{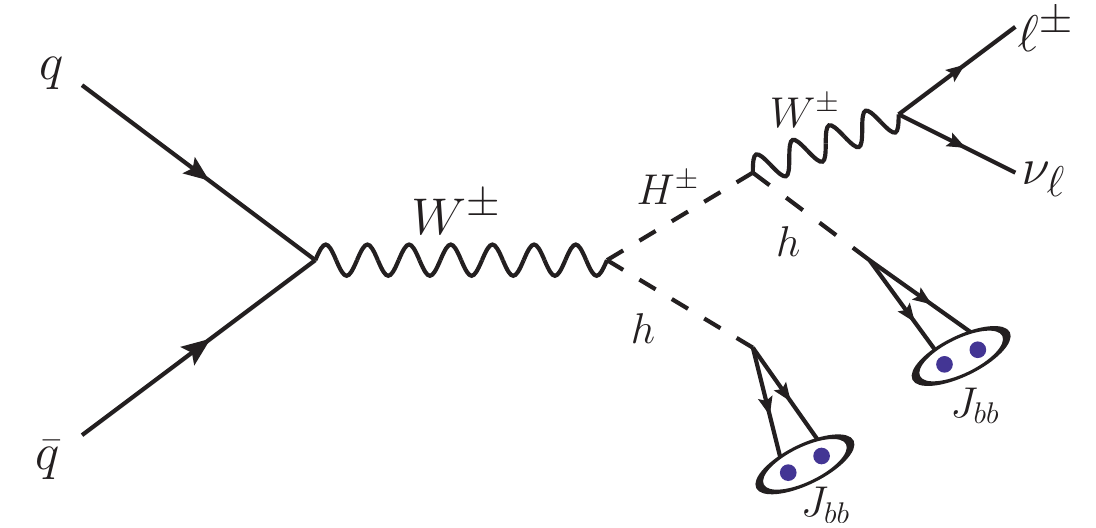}}}
		\caption{Representative Feynman diagrams for the dominant associated-production signal processes leading to final states with two fat-jets ($J_{bb}$), each containing two $b$-subjets, and one or two leptons. The additional heavy scalar pair-production channels, $AH^\pm$ and $H^+H^-$, are included in the numerical analysis but are not shown separately.}\label{fig:FeynDiag}
	\end{center}
\end{figure}

We require the electroweak gauge bosons appearing in the decay chain to decay leptonically in order to suppress QCD backgrounds and define the one- and two-lepton signal regions. The first two channels, $pp\to hA$ and $pp\to hH^\pm$, correspond to associated production of a light scalar with a heavy scalar and constitute the dominant contributions. The last two channels, $pp\to AH^\pm$ and $pp\to H^+H^-$, correspond to heavy-scalar pair production. Although their production rates are smaller, they lead to the same multi-$h$ topology after the decays $A\to hZ$ and $H^\pm\to hW^\pm$, and are therefore included in the numerical analysis. For illustration, the Feynman diagrams for the two dominant associated-production modes are shown in \autoref{fig:FeynDiag}. The diagrams for the heavy scalar pair-production modes are not shown separately, as they follow directly from the same gauge mediated production and decay vertices.



The production is governed by electroweak gauge couplings and is therefore insensitive to the suppressed fermionic interactions in the Type-I 2HDM. This makes these channels particularly relevant in the large $\tan\beta$ regime. In the last four columns of \autoref{Tab:BPs}, we present the production cross-sections for the signal processes corresponding to all benchmark points. The cross-sections are computed at leading order using {\tt MadGraph5\_aMC@NLO}~\cite{Alwall:2011uj,Alwall:2014hca} and rescaled by a constant $K$-factor of 1.34 to account for higher-order QCD corrections. As expected, the cross-sections decrease with increasing scalar masses due to phase space suppression. However, for larger mass hierarchies, the resulting light scalar becomes increasingly boosted, which enhances the efficiency of fat-jet reconstruction and partially compensates for the reduced production rate.


Based on the decay topology discussed above, the signal is characterized by 
the presence of boosted light scalars reconstructed as fat-jets ($J_{bb}$), 
along with leptons originating from the decay of the associated gauge boson. 
Depending on the decay modes of the vector bosons and reconstructed multiplicity of the fat-jets coming from the boosted scalars, we define the following signal regions:
\begin{eqnarray}
\label{Eq:FinalState}
	1\ell + 1\,J_{bb}, \qquad
	1\ell + 2\,J_{bb}, \nn\\
	2\ell + 1\,J_{bb}, \qquad
	2\ell + 2\,J_{bb}.
\end{eqnarray}
Here, $J_{bb}$ denotes a fat-jet containing two $b$-subjets, originating from  the decay $h \to b\bar{b}$. Selection of $J_{bb}$ candidate are discussed later in this section. The light leptons ($e$ or $\mu$) arise from the leptonic decay of the associated $Z$ or $W^\pm$ boson.


The final states with two fat-jets are expected to provide the cleanest signal topology, as both light scalars are reconstructed in the boosted regime. The single fat-jet categories account for cases where one of the scalars fails to satisfy the fat-jet reconstruction criteria or falls outside the detector acceptance. The dilepton final states receive dominant contributions from processes containing a $Z$ boson in the decay chain, namely $pp\to hA$ and $pp\to AH^\pm$, where the $Z$ boson can decay leptonically. The $pp\to H^+H^-$ channel can also populate the dilepton categories when both $W$ bosons decay leptonically, although its relative contribution is expected to be reduced once the dilepton invariant-mass requirement around the $Z$ pole is imposed. In contrast, the single-lepton final states are mainly driven by channels containing a leptonically decaying $W^\pm$ boson, with the dominant contribution coming from $pp\to hH^\pm$ due to its larger production cross-section, as shown in \autoref{Tab:BPs}. Additionally, a fraction of events originating from $Z$ boson decays can also populate the single-lepton category when one of the leptons is not reconstructed. 


\subsection{Background Processes}\label{sec:backgrounds}

Four different signal regions with the final states defined in \autoref{Eq:FinalState} receive contributions from several Standard Model processes that can mimic the boosted double-$b$ signature. The dominant background processes considered in this analysis are:
\begin{itemize}
	
	\item $\mathbf{Z\,H:}$  
	This process closely resembles the signal topology with a SM Higgs boson ($M_H = 125$ GeV) decaying to $b\bar{b}$ and a $Z$ boson decaying leptonically.
	
	\item $\mathbf{W^\pm\,H:}$ 
	Similar to $ZH$, this process contributes to the single-lepton final states, 
	with the Higgs decaying to $b\bar{b}$ and the $W^\pm$ boson decaying leptonically.
	
	\item $\mathbf{t\bar{t} + jets:}$ 
	This is one of the dominant backgrounds in most signal regions due to its large production cross-section. The decay of top quarks produces multiple $b$-jets and $W^\pm$ bosons, which, when decaying leptonically, can mimic the signal, especially when the resulting jets are clustered into fat-jets containing two $b$-subjets.
	
	\item $\mathbf{t\bar{t}V:}$  
	Although subleading in rate, particularly in the two-$J_{bb}$ final states, this process contributes significantly to multi-lepton signatures due to the presence of an additional gauge boson ($V = W^\pm, Z$) in the decay chain. However, its production cross-section is considerably smaller than that of $t\bar{t}$, owing to the associated production of an extra heavy particle, which leads to phase space suppression.
	
	\item $\mathbf{t\bar{t}\,H:}$  
	This background closely resembles the signal topology, as the SM Higgs boson can 
	decay into $b\bar{b}$, producing multiple $b$-jets and leptons. When the Higgs boson is moderately boosted, its decay products may be reconstructed as a double-$b$ fat-jet ($J_{bb}$), directly mimicking the signal signature. The leptonic decay of the accompanying top-quark contribute to the lepton multiplicity, enhancing its relevance as a probable background.
	
	\item \textbf{Single top:}  
	Single top production via the ~$tb$, $tW$, and $t+jets$ channels contributes to the background through leptonic top decays and associated $b$-jets. These processes can mimic the signal when additional jets are misidentified as $b$-jets and  clustered into fat-jets. Although sub-leading compared to $t\bar{t}$, it typically constitutes one of the next-to-leading background contributions across the signal regions.
	
	\item $\mathbf{Vb\bar{b}:}$  
	This is one of the most important irreducible backgrounds, as it directly produces a $b\bar{b}$ pair in association with a gauge boson. When the two $b$-jets are sufficiently collimated, they can be reconstructed as a fat-jet, closely mimicking the signal signature. The leptonic decay of the gauge boson further leads to final states with one or two leptons and one or two $J_{bb}$ candidates.
	
	\item $\mathbf{V+ \rm{\textbf{jets}}:}$  
	This background has a large production cross-section and contributes significantly to the signal regions. While it is dominated by light-flavor jets, it can also contain heavy-flavor components arising from gluon splitting or matrix-element level production. It contributes when light jets are misidentified as $b$-jets and subsequently clustered into fat-jets, thereby mimicking the signal topology. The leptonic decay of the gauge boson leads to final states with one or two leptons.
	
	\item $\mathbf{VV+ \rm{\textbf{jets}}:}$  
	Compared to the $V+ \rm{jets}$ background, the $VV+ \rm{jets}$ processes are comparatively smaller but benefit from the presence of multiple leptons originating from gauge boson decays. In the presence of additional jets, these processes can also give rise to fat-jets containing two $b$-subjets, leading to final states that closely resemble the signal topology.
	
\end{itemize}

The signal and background samples are generated at leading order and normalized using process dependent inclusive $K$-factors to approximately account for higher-order QCD corrections. This treatment captures the overall normalization effect but does not include possible differential higher-order corrections that may affect particular phase-space regions.

The expected background yields  after baseline selection ($\mathcal{C}_0$), mentioned in the next sub-section, for the different final states at the LHC with $\sqrt{s}=14~\rm{TeV}$ and an integrated luminosity of $3000~\rm{fb}^{-1}$ are summarised in \autoref{Tab:FS_BG}. Across all channels, the background is predominantly driven by the $t\bar{t}+ \rm{jets}$ and $V+\rm{jets}$ processes, reflecting their large production cross-sections. Subleading but significant contributions arise from single top production, followed by $Vb\bar{b}$ and $VV+\text{jets}$ processes. This hierarchy remains consistent across both single- and dilepton final states, although their relative importance varies with the object multiplicity.


\begin{table*}[t]
	\renewcommand{\arraystretch}{1.5}
	\centering
	\hspace*{-1.0cm}
	\scalebox{0.75}
	{\begin{tabular}{|c||c|c|c|c|c|c|c|c|c||c|}
			\hline
			\multirow{2}{*}{Final States} &\multicolumn{9}{c||}{Background events from SM processes contributing in each signal region} & \multirow{1.5}{*}{Total} \\ 
			\cline{2-10}
			& $Zh$  & $W^\pm h$ & $t \bar{t}$ + jets & $t\bar{t}V$ & $t\bar{t}h $ & Single $t$ & $V b \bar{b}$ & $V$ + jets & $VV$ + jets & \multirow{0.9}{*}{Events}  \\ 
			\hline \hline
			$1 \ell + 1\,J_{bb}$ & 1142.53 & 6414.96 & $9.87 \times 10^6$ & $2.97 \times 10^4$ & $2.99 \times 10^4$ & $1.06 \times 10^6$ & $4.02 \times 10^5$ & $1.29 \times 10^7$ & $3.55 \times 10^5$ & $2.47 \times 10^7$ \\ 
			\hline
			$1 \ell + 2\,J_{bb}$  & 2.94 & 33.10 & $5.40 \times 10^4$ & 531.52 & 1099.15 & 3637.74 & 1763.01 & $3.16 \times 10^4$ & 2202.21 & $9.48 \times 10^4$ \\ 
			\hline \hline
			$2 \ell + 1\,J_{bb}$ & 726.04 & 23.65 & $1.58 \times 10^5$ & 5782.71 & 1091.87 & $4.68 \times 10^4$ & $8.15 \times 10^4$ & $5.25 \times 10^5$ & $4.86 \times 10^4$  & $8.68 \times 10^5$  \\  
			\hline
			$2 \ell + 2\,J_{bb}$ & 2.83 & 0.0 & 610.75 & 63.66 & 35.61 & 0.0 & 272.51 & 2167.56 & 244.69 & 3397.61 \\ 
			\hline
	\end{tabular}}
	\caption{Expected number of background events after the baseline $\mathcal{C}_0$ cut (mentioned in the next sub-section) for different final states at the LHC with $\sqrt{s}=14~\rm{TeV}$ and an integrated luminosity of 
	$3000~\rm{fb}^{-1}$. Individual contributions from various Standard Model processes are shown, along with the total background yield. All results include appropriate $K$-factors to accommodate higher order corrections.}  \label{Tab:FS_BG}
\end{table*}

The total background reduces considerably with increasing lepton and fat-jet multiplicity. The $2\ell + 2J_{bb}$ final state, driven by the simultaneous requirement of two isolated leptons and two $b$-tagged fat-jets, is the cleanest channel at the level of raw event yields.  However, a lower background rate does not necessarily translate into optimal sensitivity. The interplay between signal efficiency and background rejection can significantly impact the final significance. We therefore examine how these features evolve after the full event selection, which we discuss in the next section.

\subsection{Fat-jet Reconstruction and Baseline Selection}
\label{sec:reco}
To study the signal and background processes, we employ a full Monte-Carlo simulation chain followed by a detailed object reconstruction strategy. The signal corresponding to the Two-Higgs Doublet Model is implemented using \texttt{FeynRules}~\cite{Alloul:2013bka}, where the model Lagrangian is defined and the corresponding Feynman rules are derived. The model is then exported in the \texttt{UFO}~\cite{Degrande:2011ua} format and interfaced with \texttt{MadGraph5\_aMC@NLO v3.4.2}~\cite{Alwall:2011uj,Alwall:2014hca} for event generation and subsequent analysis. Standard Model background processes are generated within the same framework, with decay modes chosen to reproduce the relevant final states under study. 
The generated parton-level events are subsequently passed to \texttt{Pythia-8.3}~\cite{Sjostrand:2006za,Sjostrand:2014zea} for parton showering and hadronization. Detector effects are simulated using \texttt{Delphes-3.5.0}~\cite{deFavereau:2013fsa}, with a modified CMS detector card. Jet clustering is performed using \texttt{FastJet}~\cite{Cacciari:2011ma} as implemented within the \texttt{Delphes} framework.

In addition to the standard reconstructed objects, large-radius jets (fat-jets, denoted as $J$) are reconstructed to capture boosted scalar decays into $b\bar{b}$. For this purpose, the Delphes configuration is extended by implementing the anti-$k_T$~\cite{Cacciari:2008gp} clustering algorithm with a radius parameter $R = 0.8$, suitable for identifying collimated two-prong topologies. We also impose a minimum transverse momentum requirement of $p_T^{\text{J}} > 100~\text{GeV}$.

We have not explicitly included the pile-up effects in our simulation, which may arise from additional interactions in nearby bunch crossings and increase the apparent jet momentum by contributing additional tracks and calorimetric energy deposits, especially for large-radius jets. Previous studies with large radius jets have shown that such an impact can be controlled by using pile-up-insensitive observables, grooming, or dedicated pile-up mitigation techniques~\cite{ATLAS:2020bbn}. Since our analysis relies on the reconstruction of large-radius double-$b$ fat-jets, a complete assessment of pile-up effects would require a dedicated experimental study with pile-up overlay, calibrated boosted-object taggers, and data-driven background estimates. Therefore, the projected sensitivities reported in this work should be interpreted as fast simulation based phenomenological estimates.


Following object reconstruction, we consider final states containing boosted scalar candidates in association with leptonic decays of gauge bosons. As a possible candidate of a boosted scalar, emphasis is placed on identifying fat-jets consistent with a two-prong $b\bar{b}$ substructure arising from $h \to b\bar{b}$ decays. The procedure for characterising such fat-jets is described below.

\paragraph{Matching of two $b$-hadron within fat-jets:}
Reconstructed fat-jets are matched to truth-level $b$-hadrons to identify candidates originating from light scalar decays. A $b$-hadron is associated with a fat-jet if it lies within $\Delta R < 0.8$ of the jet axis. A fat-jet is required to contain at least two such $b$-hadrons. For fat-jets with multiple associated $b$-hadrons, the pairwise angular separation $\Delta R_{bb}$ between all $b$-hadron pairs is computed. A fat-jet is identified as a $J_{bb}$ candidate if at least one pair satisfies $\Delta R_{bb} > 0.3$, ensuring a resolved two-prong substructure consistent with a boosted $h \to b\bar{b}$ decay. In addition, a tagging efficiency of $75\%$ is assumed for identifying a $J_{bb}$ candidate, corresponding to a fat-jet containing two $b$-subjets \cite{CMS:2024ddc}.

\paragraph{Trigger emulation:}
Since the signal regions considered in this work contain one or two prompt charged leptons arising from the leptonic decay of the associated electroweak gauge boson, the events can be selected using standard lepton-based triggers. In order to account for this experimental requirement, we include a simplified trigger emulation at the analysis level. For the single lepton signal regions, we require the reconstructed lepton to satisfy $p_T^\ell>35~{\rm GeV}$. For the dilepton signal regions, the two leptons are ordered in transverse momentum and are required to satisfy $p_T^{\ell_1}>25~{\rm GeV}$ and $p_T^{\ell_2}>18~{\rm GeV}$. We do not distinguish between electron and muon triggers in this phenomenological analysis; instead, these flavour-independent thresholds are chosen as conservative offline requirements, guided by typical CMS single- and dilepton trigger thresholds used in analyses with charged leptons and boosted large radius jets~\cite{CMS:2026hwa,CMS:2024aqx}. The trigger response is approximated by a step function and is taken to be fully efficient above these thresholds.

\paragraph{Baseline event selection criteria ($\mathcal{C}_0$) :}
Having defined the boosted double-$b$ fat-jet candidate, we now impose the baseline event selection criteria $\mathcal{C}_0$. The purpose of this selection is to retain events with a leptonic trigger handle from the associated electroweak gauge boson, together with at least one reconstructed boosted light scalar candidate. Events are required to satisfy the following conditions:
\begin{itemize}
	\item The event must contain either one or two isolated charged leptons within $|\eta_\ell|<2.5$. For the single lepton categories, the reconstructed lepton is required to satisfy $p_T^\ell > 35~\text{GeV}$. For the dilepton categories, the two leptons are ordered in transverse momentum and are required to satisfy $p_T^{\ell_1} > 25~\text{GeV}$ and $p_T^{\ell_2} > 18~\text{GeV}$. These requirements emulate the corresponding single lepton and dilepton trigger selections.
    \item The event must contain at least one fat-jet identified as a $J_{bb}$ candidate, according to the matching and tagging prescription described above.
\end{itemize}
Events passing these baseline requirements are classified into single lepton and dilepton categories according to the reconstructed lepton multiplicity. They are further separated into one- and two-$J_{bb}$ categories depending on the number of reconstructed $J_{bb}$ candidates, thereby defining the four signal regions listed in \autoref{Eq:FinalState}.



\subsection{Event Selection Criteria}\label{sec:cuts}
 We begin with events satisfying the baseline selection ($\mathcal{C}_0$) defined in \autoref{sec:reco}. Subsequently, a series of kinematic and topological cuts are imposed to suppress the SM backgrounds and enhance the signal significance. The full set of cuts employed in the analysis is summarized below:

\begin{itemize}
	
	\item {\bf Dilepton invariant mass} ($\mathcal{C}_1$):  
	We require the invariant mass of the dilepton system to be consistent with 
	the $Z$ boson mass,
	\begin{equation}
		\left| M_{\ell\ell} - M_Z \right| \leq 10~\text{GeV} \, .
	\end{equation}
	This requirement is applied only to events containing two reconstructed leptons.
	
	\item {\bf Fat-jet mass window} ($\mathcal{C}_2$):  
	At least one reconstructed fat-jet ($J_{bb}$) is required to have a mass consistent with 
	the light scalar mass,
	\begin{equation}
		\left| M_{J_{bb}} - M_{h} \right| \leq 15~\text{GeV} \, .
	\end{equation}
	This selection explicitly depends on the assumed value of the light scalar mass and hence constitutes a model-dependent cut; it is therefore not applied in the model-independent analysis discussed later.
    
	\item {\bf Fat-jet mass asymmetry} ($\mathcal{C}_3$):  
	For events containing two fat-jets, we impose a mass asymmetry condition to 
	ensure that both fat-jets are consistent with originating from identical 
	parent particles,
	\begin{equation}
		\alpha = \frac{|M^{(1)}_{J_{bb}} - M^{(2)}_{J_{bb}}|}{M^{(1)}_{J_{bb}} + M^{(2)}_{J_{bb}}} \leq 0.15 \, ,
	\end{equation}
	where $M^{(1)}_{J_{bb}}$ and $M^{(2)}_{J_{bb}}$ denote the masses of the two reconstructed fat-jets 
	($J_{bb}$). This model independent cut is particularly effective in suppressing background 	processes where the jet masses arise from combinatorial or non-resonant sources.
	
\end{itemize}


\section{Results-I: Benchmark Illustration and Parameter-space Reach}\label{sec:5-result}

Having established the reconstruction strategy and the relevant background contributions, we now evaluate the signal sensitivity and present the results of our analysis. All cut-flow tables and sensitivity projections shown below include the trigger emulation requirements defined in \autoref{sec:reco} as part of the baseline selection $\mathcal{C}_0$. The quoted significances should be understood within the assumptions of the fast-simulation framework described above, in particular the use of LO event samples normalized with inclusive $K$-factors and the absence of explicit pile-up overlay. We perform the analysis for all four signal regions defined in \autoref{Eq:FinalState}. Among these, the $1\ell+2J_{bb}$ channel provides the highest sensitivity. We therefore present its cut-flow in detail in the main text, while the results for the remaining channels are summarized below and their full cut-flow tables are provided in \autoref{app:cutflow}.

To quantify the sensitivity, we use the statistical-only Asimov significance \cite{Cowan:2010js, Kumar:2015tna},
\begin{equation}
	Z_{\rm stat} =
	\sqrt{
		2\left[
		(s+b)\ln\left(1+\frac{s}{b}\right)-s
		\right]
	},
	\label{eq:significance_stat}
\end{equation}
where $s$ and $b$ denote the signal and background yields after the corresponding event selection. In the benchmark cut-flow tables and main reach projections presented below, the quoted significance, expressed in units of $\sigma$, corresponds to $Z_{\rm stat}$ unless stated otherwise.

The results for the $1\ell+2J_{bb}$ final state at the LHC with $\sqrt{s}=14~\text{TeV}$ and an integrated luminosity of 3000~\fbi are shown in \autoref{Tab:FS2_sig}. At the baseline level, $\mathcal{C}_0$, the background remains sizable. The fat-jet mass-window requirement, $\mathcal{C}_2$, suppresses the background further while retaining a large fraction of the signal events. A substantial improvement is obtained after imposing the fat-jet mass-asymmetry requirement, $\mathcal{C}_3$, which selects events where the two reconstructed fat-jets are compatible with originating from identical parent scalars. This cut is particularly effective in reducing combinatorial and non-resonant background contributions.

\begin{table*}[t]
	\renewcommand{\arraystretch}{1.25}
	\centering
	\scalebox{0.95}
	{\begin{tabular}{|c||c|c|c|c|c|}
			\hline
			\multicolumn{6}{|c|}{Final State: $1 \ell + 2 J_{bb}$ }  \\ 
			\hline
			\multirow{2}{*}{Benchmark Points} & \multicolumn{5}{c|}{Cut flow}  \\ 
			\cline{2-6}
			& Modes  & $\mathcal{C}_0$  & $\mathcal{C}_2 $ & $\mathcal{C}_3 $  & Significance ($\sigma$)  \\ 
			\hline
			\multirow{5}{*}{bp30A}  & $h A$ & 409.3 & 358.4 & 117.1 & \multirow{5}{*}{13.82}   \\ 
			& $h H^\pm$ & 1907.1 & 1791.2 & 858.4 &    \\
			& $A H^\pm$ & 1322.1 & 712.0 & 139.2 &    \\
			& $H^+ H^-$ & 845.6 & 514.3 & 117.0 &    \\
			& BG & $9.48 \times 10^4$ & $5.61 \times 10^4$ & 7547.3 &    \\
			\hline
			\multirow{5}{*}{bp30B}  & $h A$ & 129.8  & 123.1 & 54.5 &  \multirow{5}{*}{6.24}  \\ 
			& $h H^\pm$ & 861.6 & 833.9 & 423.8 &    \\
			& $A H^\pm$ & 185.8 & 145.5 & 40.8 &    \\
			& $H^+ H^-$ & 111.2 & 92.2 & 29.1 &    \\
			& BG & $9.48 \times 10^4$ & $5.61 \times 10^4$ & 7547.3 &   \\
			\hline\hline
			\multirow{5}{*}{bp50A}  & $h A$ & 313.6  & 286.8 & 156.3  &   \multirow{5}{*}{14.13}  \\
			& $h H^\pm$ & 2019.0 & 1917.1 & 1189.9 &    \\
			& $A H^\pm$ & 586.8 & 452.3 & 192.0 &    \\
			& $H^+ H^-$ & 413.2 & 333.6 & 154.1 &    \\
			& BG & $9.48 \times 10^4$  & $6.68 \times 10^4$ & $1.38 \times 10^4$ &    \\
			\hline
			\multirow{5}{*}{bp50B}  & $h A$ & 155.2 & 142.5 & 74.5 &   \multirow{5}{*}{6.95}  \\ 
			& $h H^\pm$ & 1126.8  & 1063.0 & 615.6 &    \\
			& $A H^\pm$ & 243.0 & 191.4 & 76.3 &    \\
			& $H^+ H^-$ & 169.8 & 139.1 & 58.2 &    \\
			& BG & $9.48 \times 10^4$ & $6.68 \times 10^4$  & $1.38 \times 10^4$ &   \\
			\hline\hline
			\multirow{5}{*}{bp70A}  & $h A$ & 143.8  & 125.2 & 76.2 &  \multirow{5}{*}{9.36}   \\
			& $h H^\pm$ & 942.4 & 860.1 & 576.8 &    \\
			& $A H^\pm$ & 288.8 & 214.9 & 108.8 &    \\
			& $H^+ H^-$ & 194.7 & 152.2 & 83.8 &    \\
			& BG &  $9.48 \times 10^4$ & $4.42 \times 10^4$  & 7877.8 &    \\
			\hline
			\multirow{5}{*}{bp70B} & $h A$ & 83.2 & 73.4 & 44.3 & \multirow{5}{*}{5.79}  \\ 
			& $h H^\pm$ & 622.5 & 565.6 & 368.8 &    \\
			& $A H^\pm$ & 143.2 & 113.0 & 60.0 &    \\
			& $H^+ H^-$ & 100.1 & 83.2 & 46.6 &    \\
			& BG & $9.48 \times 10^4$  & $4.42 \times 10^4$  & 7877.8 &   \\
			\hline
	\end{tabular}}
	\caption{Event yields and signal significance for the $1\ell + 2J_{bb}$ final 
		state at the LHC with $\sqrt{s}=14~\rm{TeV}$ and an integrated luminosity of  3000 \fbi. The contributions from $hA$, $hH^\pm$, $A H^\pm$ and $H^+ H^-$ signal processes, as well as the total background (BG), are shown after the baseline selection cut ($\mathcal{C}_0$), the fat-jet mass window cut ($\mathcal{C}_2$) and the fat-jet mass asymmetry cut ($\mathcal{C}_3$). The resulting statistical significance $Z_{\rm stat}$ is reported for each benchmark point.}  \label{Tab:FS2_sig}
\end{table*}
%

All four electroweak production modes, $hA$, $hH^\pm$, $AH^\pm$ and $H^+H^-$, contribute to the $1\ell+2J_{bb}$ signal region. Among them, the associated production channel $hH^\pm$ gives the dominant contribution, mainly because of its larger production cross-section and its direct connection to the single lepton topology through the leptonic decay of the $W^\pm$ boson. The $hA$ channel also contributes, particularly when one lepton from the leptonic $Z$ decay is not reconstructed or falls outside the acceptance. The heavy scalar pair production modes, $AH^\pm$ and $H^+H^-$, provide additional but subleading contributions. Although their inclusive production rates are smaller, their contribution after the full selection is non-negligible for several benchmark points and is therefore included in the total signal yield and in the significance estimate. The requirement of two reconstructed $J_{bb}$ candidates enhances sensitivity to events in which both light scalars are produced in the boosted regime. This leads to a significant improvement in the signal significance for several benchmark points. In particular, the bp30A and bp50A benchmarks reach statistical significances of more than $13\sigma$ at 3000~\fbi, while the other benchmark points yield significances above the $5\sigma$ level. For heavier scalar masses, the light scalar becomes more boosted; however, the corresponding decrease in the production cross-section reduces the total event yield and hence the overall significance. Nevertheless, the $1\ell+2J_{bb}$ channel demonstrates strong sensitivity, highlighting the advantage of reconstructing both light scalars in the boosted regime and identifying this final state as the most promising discovery channel in the present analysis.

To assess the robustness of these projections against background systematics, we also evaluate the profile-likelihood Asimov significance including a Gaussian background uncertainty \cite{Cowan:2010js, Kumar:2015tna},
\begin{equation}
	Z_{\rm sys} = \left[2\left\{(s+b)\ln \left[	\frac{(s+b)(b+\sigma_b^2)} {b^2+(s+b)\sigma_b^2} \right]
	- \frac{b^2}{\sigma_b^2} \ln \left[	1+\frac{\sigma_b^2 s}{b(b+\sigma_b^2)} \right] \right\}	\right]^{1/2},
	\label{eq:significance_sys}
\end{equation}
where $\sigma_b=\delta_b b$ parametrizes the absolute background uncertainty. The parameter $\delta_b$ can be interpreted as an effective fractional
uncertainty on the total background yield after the full event selection. In the limit $\sigma_b\to0$, \autoref{eq:significance_sys} reduces to \autoref{eq:significance_stat}.

As expected, the impact of a flat normalization uncertainty on the total background can be sizable in the present cut-based analysis. For illustration, in the most sensitive $1\ell+2J_{bb}$ channel, the statistical-only significance of the best benchmark point (bp50A) is reduced from about $14.13\sigma$ to about $5.44\sigma$ when a representative $2\%$ uncertainty is assigned to the total background normalization. This indicates that the ultimate sensitivity will depend on how well the relevant backgrounds can be constrained in a dedicated experimental analysis. In practice, control regions, sidebands in the fat-jet mass distribution, shape information, and data-driven background estimates can reduce the effective background uncertainty. In addition, a multivariate analysis exploiting the full event kinematics and jet-substructure information could improve the signal-to-background ratio and mitigate the impact of residual systematics. Such an optimization is beyond the scope of the present work.

The remaining three final states exhibit reduced sensitivity for different reasons. The $1\ell+1J_{bb}$ channel receives contributions from the same four signal-production modes, but is limited by the large residual background and by the absence of the mass-asymmetry requirement $\mathcal{C}_3$. The dilepton channels, $2\ell+1J_{bb}$ and $2\ell+2J_{bb}$, provide cleaner final states through the reconstruction of the leptonically decaying $Z$ boson, but suffer from smaller signal yields and do not receive contributions from the dominant $hH^\pm$ production mode. In these channels, the $hA$ and $AH^\pm$ modes provide the leading signal contributions, since they contain a $Z$ boson that can decay to a same-flavour dilepton pair and satisfy the invariant-mass requirement $\mathcal{C}_1$. The $H^+H^-$ contribution is comparatively smaller because the dileptons arise from two separate $W$ boson decays rather than from an on-shell $Z$ boson, and therefore are strongly suppressed by the $Z$-mass requirement $\mathcal{C}_1$. Although the $2\ell+2J_{bb}$ channel achieves strong background suppression, the simultaneous requirement of two leptons and two reconstructed $J_{bb}$ candidates significantly reduces the signal statistics. A detailed cut-flow analysis for these subleading channels is given in \autoref{app:cutflow}.


\subsection{Discovery and Exclusion Reach}

After establishing the effectiveness of the proposed reconstruction and selection strategy using representative benchmark points, we proceed to determine the sensitivity of the model through a continuous scan of the relevant parameter space. The benchmark points serve only to illustrate the event selection and cut-flow behaviour; the final collider reach is not extrapolated from them. For each representative light-scalar mass, $M_h = 30$, $50$, and $70~\mathrm{GeV}$, we scan the heavy pseudoscalar mass over the range $M_A = 150$--$650~\mathrm{GeV}$ and evaluate the signal significance at each point. We further extend the scan to the $(M_A,\tan\beta)$ plane to determine the discovery and exclusion reach across the viable parameter space of the scenario under consideration.

\begin{figure}[t]
	\begin{center}
		\mbox{\subfigure[]{\includegraphics[width=0.48\linewidth,angle=0]{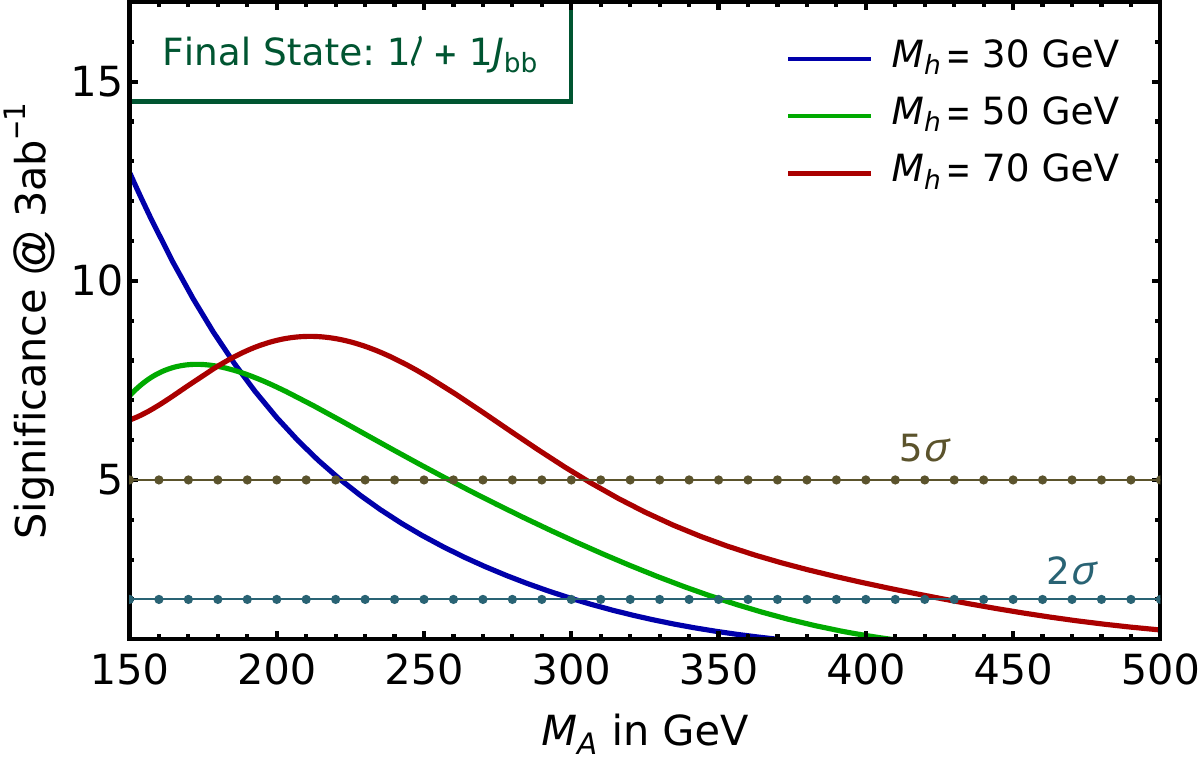}}\quad \quad
			\subfigure[]{\includegraphics[width=0.48\linewidth,angle=0]{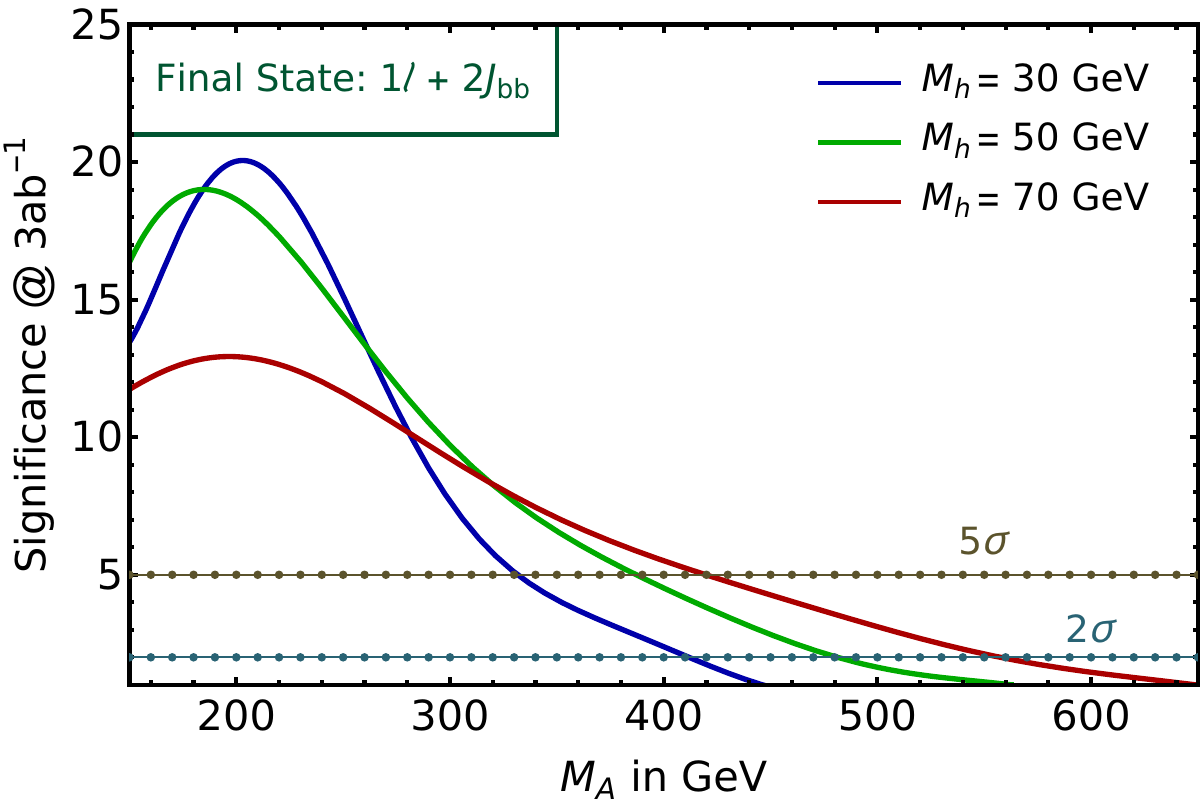}}}
		\mbox{\subfigure[]{\includegraphics[width=0.48\linewidth,angle=0]{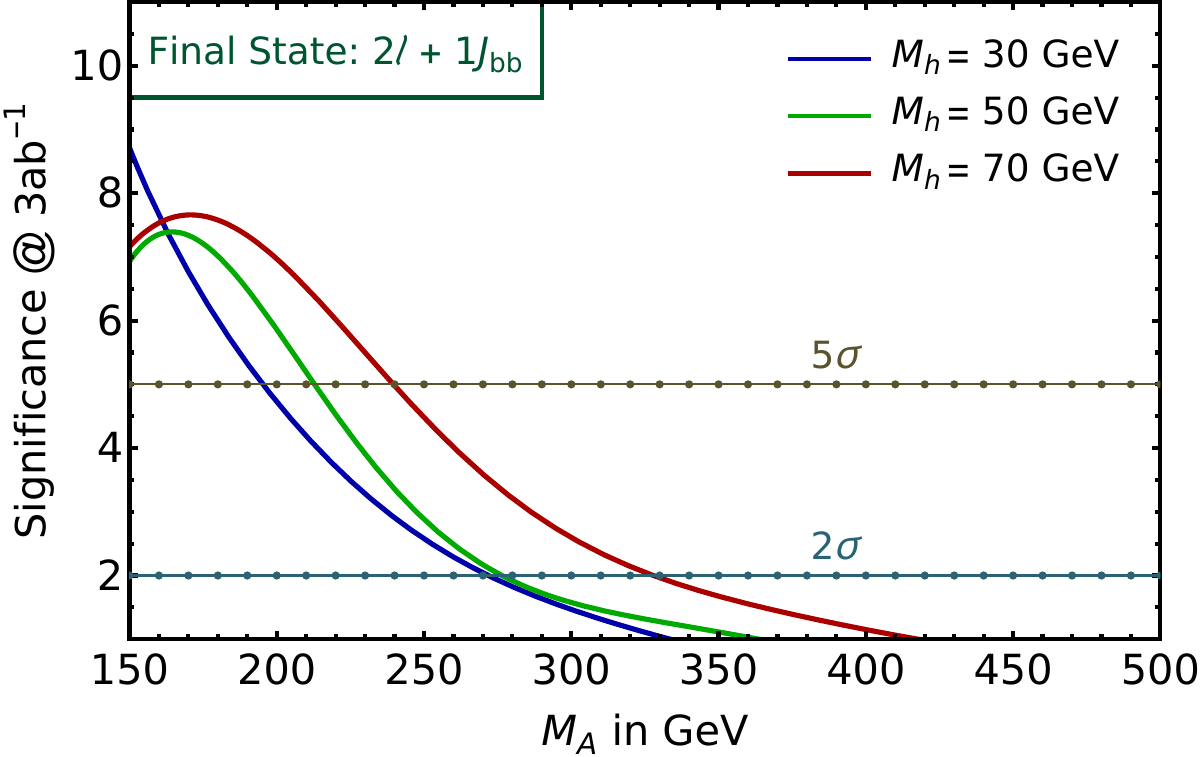}}\quad \quad
			\subfigure[]{\includegraphics[width=0.48\linewidth,angle=0]{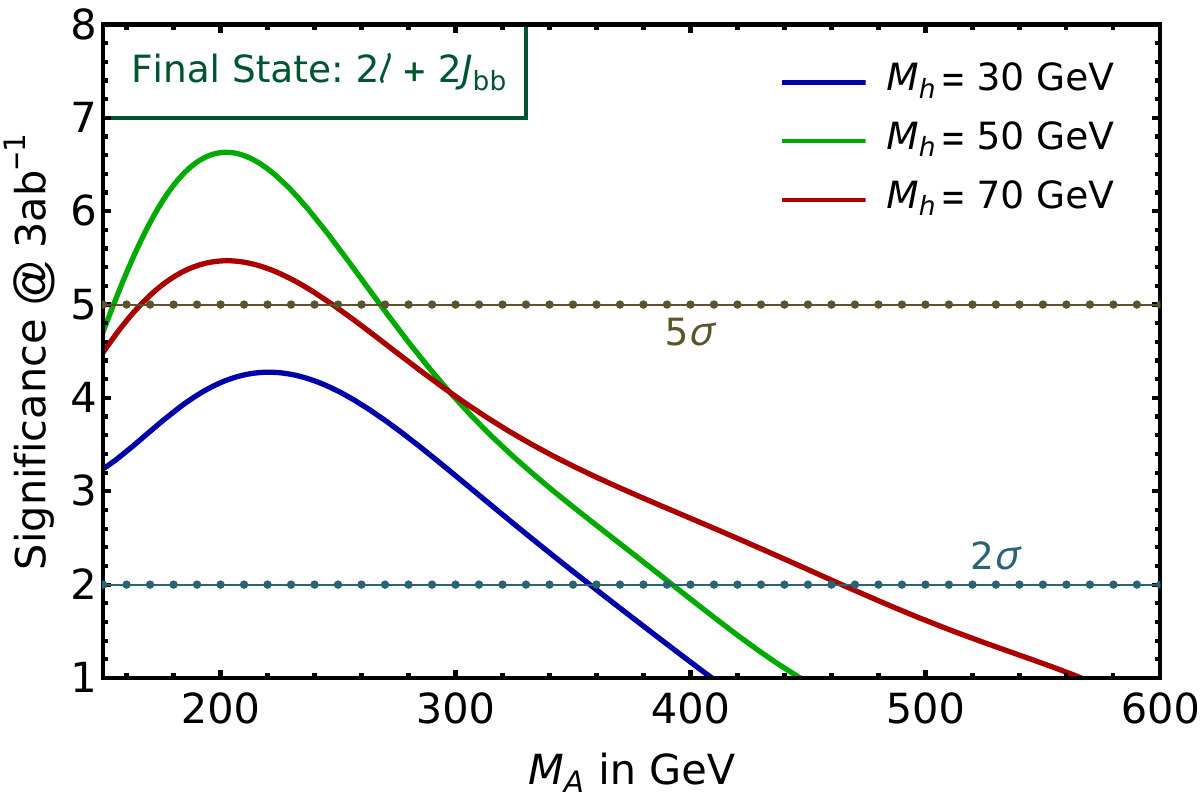}}}
		\caption{Signal significance as a function of $M_{A}$ for different final states at the LHC with $\sqrt{s}=14~\rm{TeV}$ and an integrated luminosity of 3000~\fbi. Panels (a)--(d) correspond to the $1\ell + 1J_{bb}$, $1\ell + 2J_{bb}$, $2\ell + 1J_{bb}$, and $2\ell + 2J_{bb}$ final states, respectively. The blue, green, and red curves represent $M_{h} = 30$, $50$, and $70~\rm{GeV}$, respectively. The horizontal lines indicate the $2\sigma$ and $5\sigma$ significance levels. The $1\ell + 2J_{bb}$ channel exhibits the highest sensitivity, while the dilepton channels, although cleaner, are limited by reduced signal statistics.} \label{fig:ReachPlot}
	\end{center}
\end{figure}


The discovery reach for the different final states at the LHC with a centre-of-mass energy of $\sqrt{s}=14~\text{TeV}$ and an integrated luminosity of 3000~\fbi is illustrated in \autoref{fig:ReachPlot}. Each panel shows the signal significance as a function of the heavy scalar mass for different choices of the light scalar mass. The blue, green, and red curves represent $M_{h} = 30$, $50$, and $70~\text{GeV}$, respectively. The mass difference between $H^\pm$ and $A$ is fixed to $5~\text{GeV}$ throughout, with $M_{H^\pm} > M_{A}$.

Several features can be inferred from the significance distributions shown in \autoref{fig:ReachPlot}. In all channels, the sensitivity decreases with increasing $M_{A}$, reflecting the suppression of the production cross-section at higher masses. In channels with two reconstructed fat-jets, a broad maximum is observed at intermediate masses. This behavior can be attributed to the interplay between the increasing boost of the light scalar, which enhances the reconstruction efficiency, and the simultaneous reduction in the production rate. 
The dependence on the light scalar mass further demonstrates the interplay between boost and reconstruction effects. Lighter scalars ($M_h = 30~\text{GeV}$) receive larger boosts at low $M_A$, but the resulting collimation of the $b$-quarks leads to merging of substructure and a loss of tagging efficiency at higher masses. In contrast, for heavier scalars ($M_h = 70~\text{GeV}$), the decay products remain better resolved within the fat-jet, allowing for more stable reconstruction and a slower degradation of sensitivity with increasing $M_A$. This leads to the observed variation across the parameter space.

\begin{figure}[t]
	\begin{center}
		\mbox{\subfigure[]{\includegraphics[width=0.55\linewidth,angle=0]{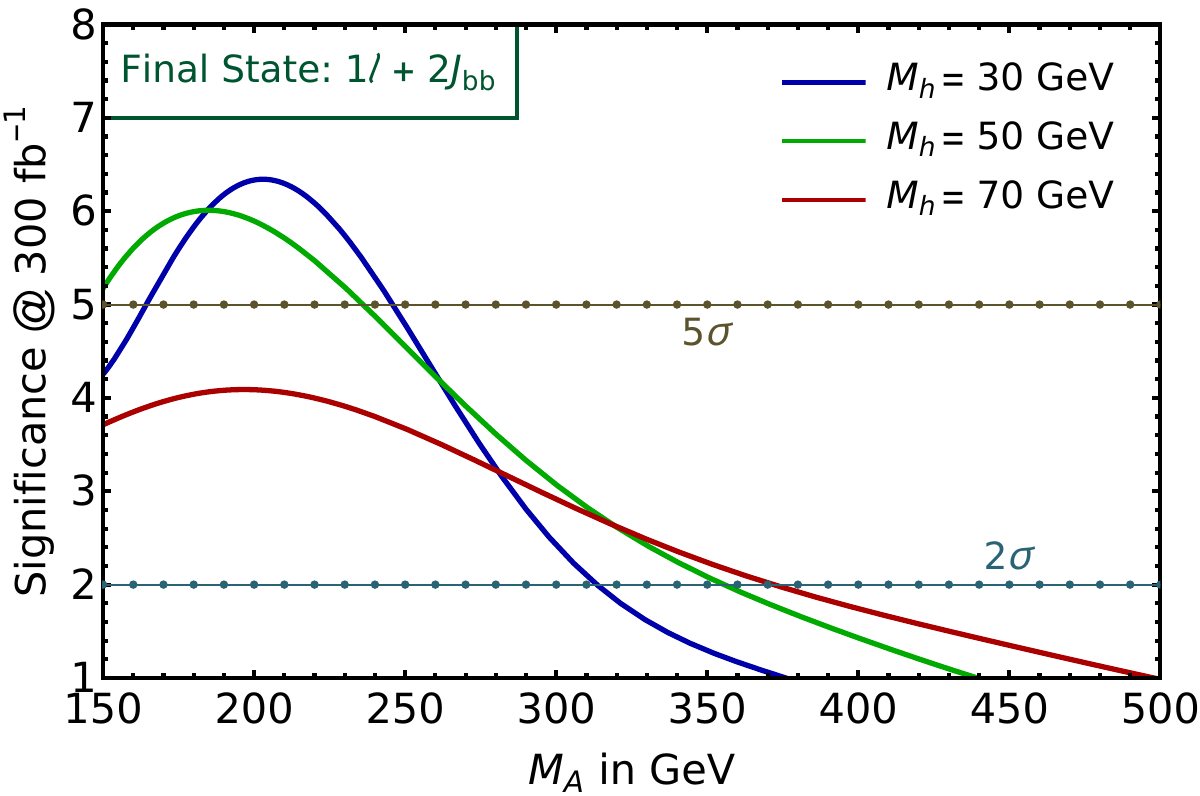}}}
		\caption{Signal significance as a function of $M_{A}$ for the $1\ell + 2J_{bb}$ final state at the LHC with $\sqrt{s}=14~\rm{TeV}$ and an integrated luminosity of 300~\fbi. The color coding and significance levels are the same as in \autoref{fig:ReachPlot}.}\label{fig:RchPlt300}
	\end{center}
\end{figure}

Consistent with the cut-flow analysis, a clear hierarchy among the channels is observed. The $1\ell + 2J_{bb}$ final state  (top--right panel) exhibits the highest sensitivity across a wide mass range, reaching significances well above $5\sigma$ in the lower mass region, with discovery reach extending up to $M_{A} \approx 330$, $390$, and $420~\text{GeV}$ for $M_{h} = 30$, $50$, and $70~\text{GeV}$, respectively. Even at higher masses, this channel maintains observable sensitivity, achieving $2\sigma$ significance up to $M_{A} \approx 560~\text{GeV}$ for $M_{h} = 70~\text{GeV}$. This represents the maximum reach of the model within the boosted double-$b$ fat-jet framework considered in this work. 
In contrast, the $1\ell + 1J_{bb}$ channel (top--left panel) shows a rapid degradation in sensitivity with increasing mass due to large residual backgrounds. For instance, for $M_{h} = 50~\text{GeV}$, the reach is limited to $M_{A} \approx 350~\text{GeV}$ at the $2\sigma$ level, compared to $M_{A} \approx 480~\text{GeV}$ in the $1\ell + 2J_{bb}$ channel. 

In the bottom panels, the dilepton channels, $2\ell + 1J_{bb}$ and $2\ell + 2J_{bb}$, exhibit a reduced reach across the parameter space, primarily due to the lower signal yield. For instance, in the $2\ell + 1J_{bb}$ channel, the $2\sigma$ sensitivity extends up to $M_{A} \sim 330~\text{GeV}$ for $M_{h} = 70~\text{GeV}$, while for the lighter benchmark points ($M_{h} = 30, 50~\text{GeV}$), the reach is limited to around $M_{A} \sim 275~\text{GeV}$. In comparison, the $2\ell + 2J_{bb}$ channel shows an improved reach, extending up to $M_{A} \approx 360$, $397$, and $462~\text{GeV}$ for $M_{h} = 30$, $50$, and $70~\text{GeV}$, respectively. Despite providing the cleanest topology, the stringent requirements on both leptons and fat-jets significantly reduce the overall signal statistics, preventing this channel from surpassing the sensitivity of the $1\ell + 2J_{bb}$ final state.

Given that the $1\ell + 2J_{bb}$ final state exhibits the highest sensitivity among all the channels considered, we further investigate its discovery prospects at a lower integrated luminosity of 300~\fbi. The corresponding signal significance is shown in \autoref{fig:RchPlt300}. While the discovery reach is naturally reduced compared to the 3000~\fbi case, the channel retains substantial sensitivity in the lower mass region. In particular, for $M_h=30$ and $50~\text{GeV}$, the signal significance exceeds the $5\sigma$ level for relatively light $A$ masses, reaching a maximum around $M_A \sim 180$--$210~\text{GeV}$. For $M_h=70~\text{GeV}$, the significance remains below the discovery threshold, but still reaches about $4\sigma$ in the low mass region.


More importantly, a significant region of the parameter space remains accessible at the exclusion level. The $2\sigma$ sensitivity extends to $M_{A} \gsim 370\text{GeV}$ for most benchmark choices, indicating that a wide range of masses can already be probed with early data. As in the high-luminosity case, the sensitivity decreases with increasing $M_{A}$ due to phase space suppression and the corresponding reduction in production cross section.
Overall, even at 300~\fbi, the $1\ell + 2J_{bb}$ channel provides a robust probe of the inverted Type-1 2HDM.

\begin{figure}[t]
	\begin{center}
		\mbox{\subfigure[]{\includegraphics[width=0.48\linewidth,angle=0]{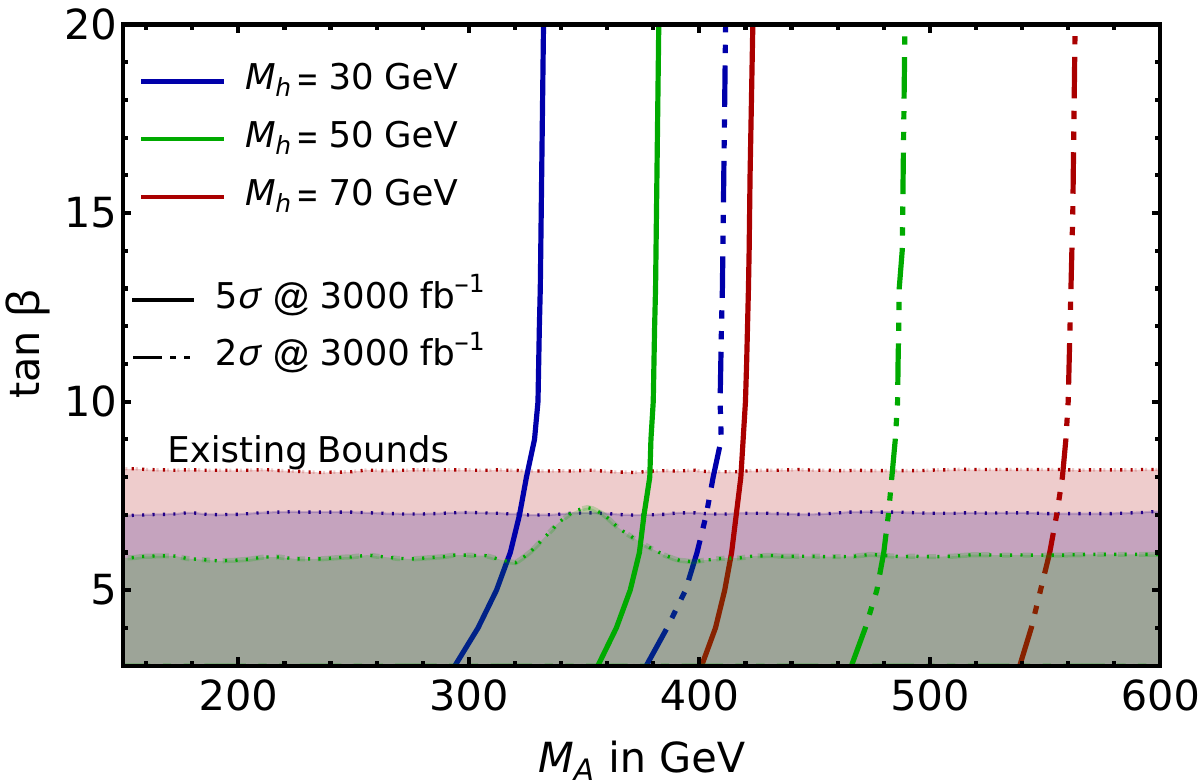}}\quad
			\subfigure[]{\includegraphics[width=0.48\linewidth,angle=0]{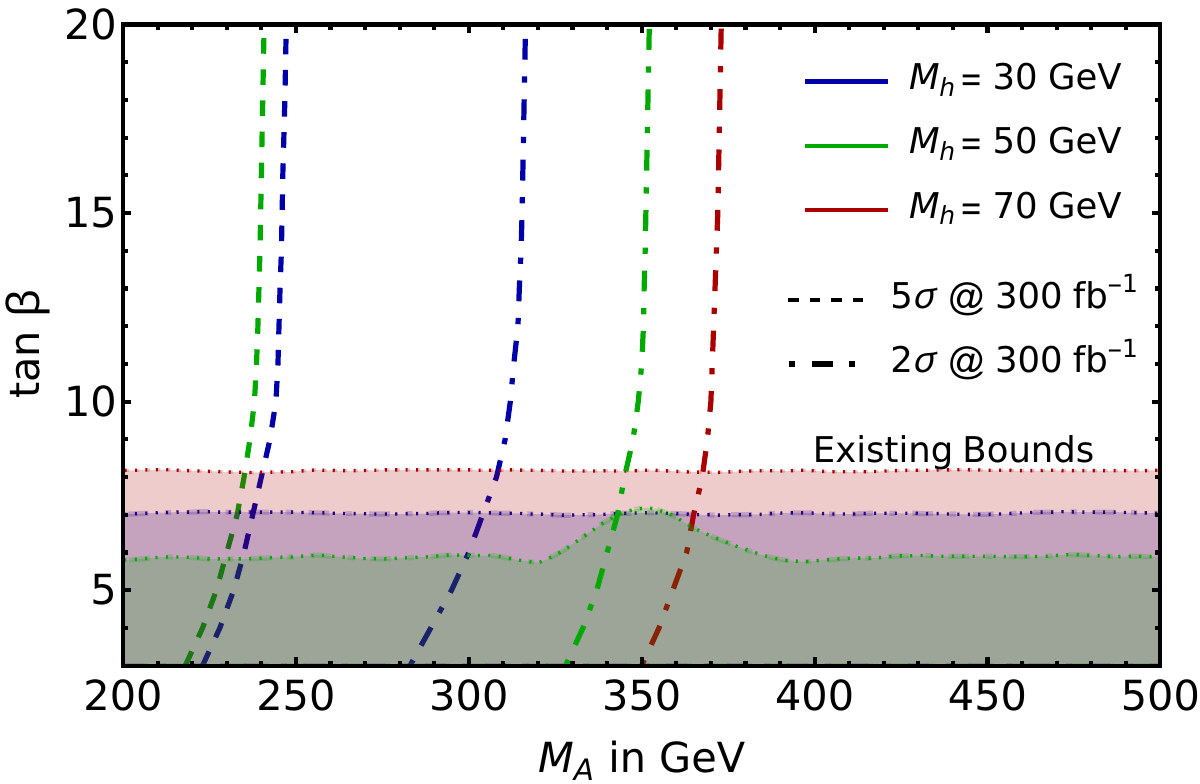}}} 
		\caption{Projected sensitivity in the $M_A$ versus $\tan\beta$ plane for the $1\ell+2J_{bb}$ final state at the LHC with $\sqrt{s}=14~\mathrm{TeV}$. Panel (a) shows the $2\sigma$ and $5\sigma$ reach for an integrated luminosity of 3000~\fbi, while panel (b) shows the corresponding reach for 300~\fbi. The blue, green, and red contours correspond to $M_h=30$, $50$, and $70~\mathrm{GeV}$, respectively. The different line styles denote the $2\sigma$ and $5\sigma$ sensitivity contours, as indicated in the legends. The shaded region represents the parameter space excluded by the existing theoretical and experimental bounds discussed in  \autoref{sec:3-bp}.} 
        \label{fig:tanb-mA}
    \end{center}  
\end{figure}

\autoref{fig:tanb-mA} shows the projected sensitivity of the boosted double-$b$ fat-jet analysis for the $1\ell + 2J_{bb}$ final state in the $M_A$--$\tan\beta$ plane, where we fix $M_{H^\pm} - M_A = 5~\text{GeV}$. The shaded region at low $\tan\beta$ corresponds to existing theoretical and experimental constraints for different light scalar masses. The solid (dot-dot-dashed) curves in panel (a) denote the $5\sigma$ ($2\sigma$) reach at 3000~\fbi, while the dashed (dot-dashed) curves in panel (b) show the $5\sigma$ ($2\sigma$) reach at 300~\fbi. The blue, green, and red curves correspond to $M_h = 30$, $50$, and $70~\text{GeV}$, respectively. A substantial region at moderate to high $\tan\beta$ remains unconstrained by existing bounds. This region can be effectively probed by the proposed analysis, with the $2\sigma$ reach at 300~\fbi~already extending beyond the currently excluded region. At 3000~\fbi, both the exclusion and discovery sensitivities improve significantly, covering a large portion of the previously unexplored parameter space.

Here, the dependence on $\tb$ enters through the mixing angle between the neutral scalars, $(\beta-\alpha)$. The production cross section, as well as the decay of the heavy scalars into a light scalar and a gauge boson, are governed by $\cos(\beta-\alpha)$. The light scalar $h$ decays to $b\bar b$ essentially independent of $\tb$. From \autoref{fig:br-Hhh}, it is evident that for relatively large $\tb$ the allowed range of $\sba$ becomes nearly constant, leading to an approximately vertical exclusion boundary. At low $\tb$, larger negative values of $\sba$ are allowed, which in turn mildly affect the reach of the proposed signal.

\section{Results-II: Model-independent Analysis}
\label{sec:model_indep}

So far, the analysis has relied on a combination of kinematic and resonance based selection criteria tailored to the specific signal topology and benchmark points as discussed in \autoref{sec:cuts}. In particular, the fat-jet mass window cut ($\mathcal{C}_2$) plays a crucial role in reconstructing the light scalar resonance. However, this requirement introduces a degree of model dependence, as it relies explicitly on the knowledge of the scalar mass. To address this limitation, we perform a complementary model-independent analysis by removing the fat-jet mass window cut ($\mathcal{C}_2$). Instead, we focus on topological observables that are less sensitive to the underlying model parameters. For final states containing two fat-jets, we retain the fat-jet mass asymmetry cut ($\mathcal{C}_3$), which exploits the presence of two objects originating from identical parent particles without requiring explicit mass reconstruction. For dilepton final states, the invariant mass cut ($\mathcal{C}_1$) is also applied to ensure consistency with $Z$ boson decays.


\begin{table*}[t]
	\renewcommand{\arraystretch}{1.5}
	\centering
	\scalebox{0.97}
	{\begin{tabular}{|c||c|c|c|c|c|c||c|}
			\hline
			\multicolumn{8}{|c|}{Final State: $1 \ell + 2 J_{bb}$ ~ with the cuts $\mathcal{C}_0$ and $\mathcal{C}_3$   }  \\
			\hline
			\multirow{2}{*}{Modes} &\multicolumn{6}{c||}{Signals} & \multirow{1.5}{*}{Total Background } \\ 
			\cline{2-7}
			& bp30A  & bp30B  & bp50A & bp50B  & bp70A & bp70B &  \multirow{0.9}{*}{Contribution}  \\ 
			\hline
			$hA$ & 147.2 & 57.3  & 166.0 & 78.3  & 82.4 & 47.6 &  \multirow{4}{*}{$1.95 \times 10^4$} \\ 
			
			$hH^\pm$  & 918.4 & 432.8 & 1220.0 & 631.9  &  606.8 & 386.0  &   \\ 
			$AH^\pm$  & 452.6 & 52.5 & 239.3 & 90.7 & 133.6 & 68.6 &   \\ 
			$H^+H^-$  & 282.7 & 34.7 & 180.4 & 66.4 & 97.9 & 51.1 &   \\ 
			\hline \hline
			Total & 1801.0  & 577.4  & 1805.8 & 867.2  & 920.7 & 553.3 &  \multicolumn{1}{c}{} \\ 
			\cline{1-7}
			Significance ($\sigma$) & 12.71 & 4.11 & 12.74 & 6.16  & 6.54 & 3.94 &  \multicolumn{1}{c}{} \\ 
			\cline{1-7}
	\end{tabular}}
	\caption{Event yields and signal significance for the $1\ell + 2J_{bb}$ final state at the LHC with the centre-of-mass energy of 14 TeV and an integrated luminosity of 3000\,\fbi, using a model-independent selection based on the baseline cut ($\mathcal{C}_0$) and the fat-jet mass asymmetry cut ($\mathcal{C}_3$), without applying the fat-jet mass window cut  ($\mathcal{C}_2$). Contributions from all the production modes ($hA$, $hH^\pm$, $AH^\pm$, and $H^+H^-$) are shown, along with the total background.}  \label{Tab:FS2ind_Sig}
\end{table*}

The resulting event yields and significances for the $1\ell + 2J_{bb}$ and $2\ell + 2J_{bb}$ final states are presented in \autoref{Tab:FS2ind_Sig} and \autoref{Tab:FS4ind_Sig}, respectively. Compared to the corresponding results obtained with the full set of cuts, a moderate reduction in signal significance is observed due to the absence of the resonance-based discrimination provided by $\mathcal{C}_2$. For instance, for the bp30B benchmark point, the significance decreases from $6.24\,\sigma$ (see \autoref{Tab:FS2_sig}) to $4.11\,\sigma$. 
Despite this reduction, the sensitivity remains substantial, particularly in the $1\ell + 2J_{bb}$ channel, where several benchmark points still achieve significances above the $5\sigma$ level. In the dilepton channel, the significance is further reduced due to the smaller signal yield, but remains non-negligible for favorable benchmark configurations. Importantly, the overall trends observed in the model-dependent analysis are preserved: the $1\ell + 2J_{bb}$ channel continues to outperform the $2\ell + 2J_{bb}$ channel.


\begin{table*}[t]
	\renewcommand{\arraystretch}{1.5}
	\centering
	\scalebox{0.97}
	{\begin{tabular}{|c||c|c|c|c|c|c||c|}
			\hline
			\multicolumn{8}{|c|}{Final State: $2 \ell + 2 J_{bb}$ ~ with the cuts $\mathcal{C}_0$, $\mathcal{C}_1$ and $\mathcal{C}_3$ }  \\
			\hline
			\multirow{2}{*}{Modes} &\multicolumn{6}{c||}{Signals} & \multirow{1.5}{*}{Total Background } \\ 
			\cline{2-7}
			& bp30A  & bp30B  & bp50A & bp50B  & bp70A & bp70B &  \multirow{0.9}{*}{Contribution}  \\ 
			\hline
			$hA$ & 48.0 & 46.4 & 112.6 & 58.7 & 53.7 & 38.8 &  \multirow{3}{*}{547.38} \\ 
			$AH^\pm$  & 43.2 & 7.1 & 27.0 & 11.9 & 15.3 & 9.1 &   \\ 
			$H^+H^-$  & 3.0 & 0.4 & 2.7 & 0.8 & 1.2 & 0.5 &   \\ 
			\hline \hline
			Total & 94.2  & 54.0  & 142.3 &  71.4 & 70.1 & 48.4 &  \multicolumn{1}{c}{} \\ 
			\cline{1-7}
			Significance ($\sigma$) &  3.92 & 2.27  & 5.84 & 2.99  & 2.94 & 2.04 &  \multicolumn{1}{c}{} \\ 
			\cline{1-7}
	\end{tabular}}
	\caption{Event yields and signal significance for the $2\ell + 2J_{bb}$ final state at the LHC with $\sqrt{s}=14~\rm{TeV}$ and an integrated luminosity of 3000\,\fbi, using a model-independent selection based on the baseline cut ($\mathcal{C}_0$), the dilepton invariant mass cut ($\mathcal{C}_1$), and the fat-jet mass asymmetry cut ($\mathcal{C}_3$), without applying the fat-jet mass window cut ($\mathcal{C}_2$). Contributions from the production modes $hA$, $AH^\pm$, and $H^+H^-$ are shown, along with the total background.}  \label{Tab:FS4ind_Sig}
\end{table*}

This model-independent strategy has important implications for experimental searches. By avoiding explicit assumptions about the scalar mass, it allows for a broader interpretation of the results in terms of generic boosted di-$b$ signatures. Such an approach can be readily implemented by experimental analyses to search for new physics scenarios that produce similar final states, without being restricted to a specific benchmark model. It therefore provides a complementary and more versatile framework for probing beyond the Standard Model physics at the LHC.  

\subsection{Discovery and Exclusion Reach}
\begin{figure}[hbt]
	\begin{center}
		\mbox{\subfigure[]{\includegraphics[width=0.48\linewidth,angle=0]{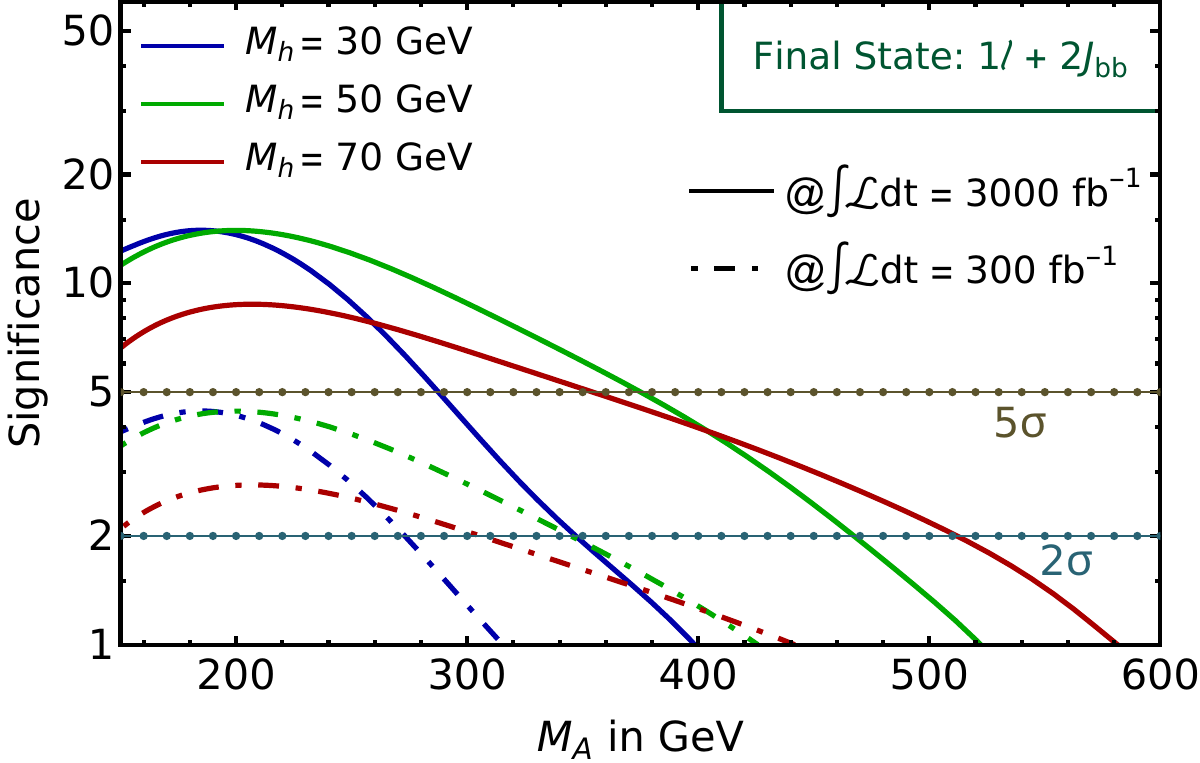}}\quad 
			\subfigure[]{\includegraphics[width=0.48\linewidth,angle=0]{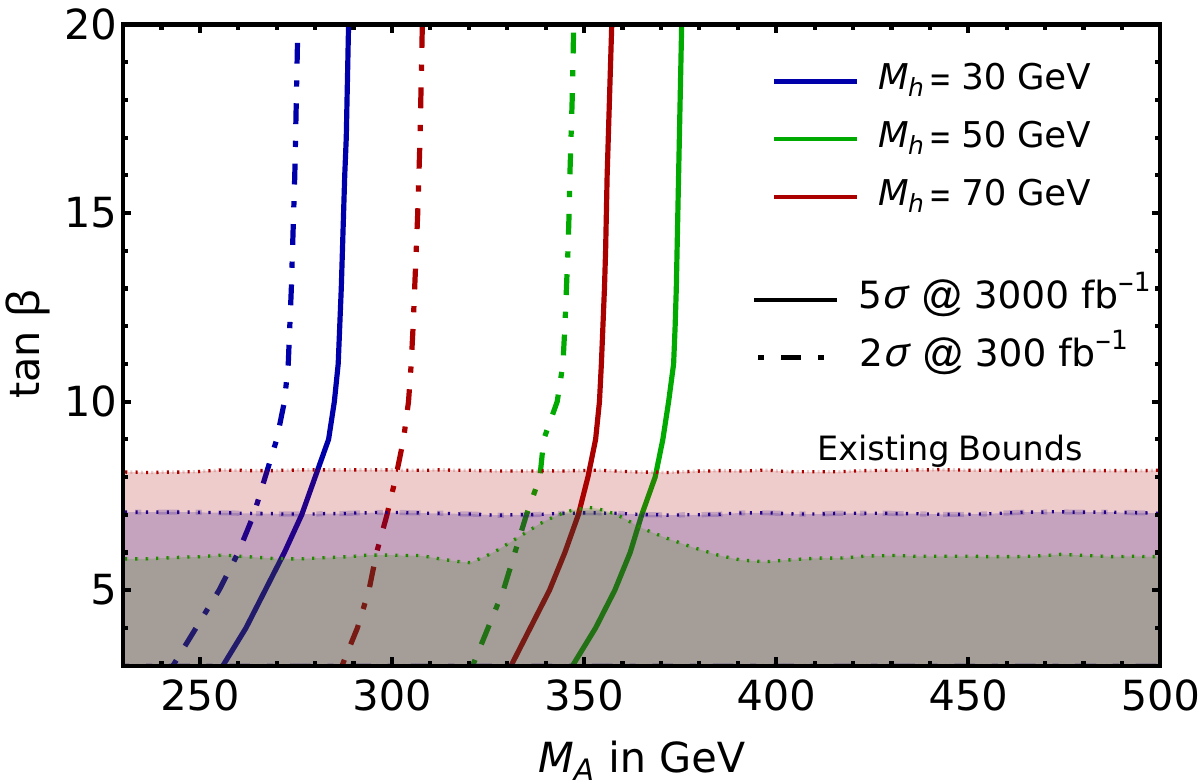}}} 
		\caption{Discovery potential for the $1\ell + 2J_{bb}$ final state in the model-independent analysis at the LHC with $\sqrt{s}=14~\rm{TeV}$. The selection criteria correspond to the model-independent cuts are defined in \autoref{Tab:FS2ind_Sig}. Panel (a) shows the signal significance as a function of $M_{A}$ for different choices of $M_{h}$ at integrated luminosities of 3000~\fbi (solid lines) and 300~\fbi (dot-dashed lines), with horizontal lines indicating the $2\sigma$ and $5\sigma$ significance levels. Panel (b) presents the corresponding sensitivity in the $M_{A}$--$\tan\beta$ plane, where the $5\sigma$ contours at 3000~\fbi (solid) and the $2\sigma$ contours at 300~\fbi (dot-dashed) are shown, along with existing bounds (shaded region). The blue, green, and red curves correspond to $M_{h} = 30$, $50$, and $70~\rm{GeV}$, respectively.} \label{fig:RchPlt_MdlInd}
	\end{center}
\end{figure}

To further quantify the discovery potential within this model-independent framework, we examine the reach of the most sensitive  $1\ell + 2J_{bb}$ final state. The corresponding signal significance as a function of $M_{A}$ at the LHC with $\sqrt{s}=14~\rm{TeV}$ and an integrated luminosity of 3000\,\fbi (solid lines) and 300\,\fbi (dot-dashed lines) are shown in \autoref{fig:RchPlt_MdlInd} (a).


A behaviour consistent with the model-dependent analysis is observed, with the sensitivity exhibiting a clear dependence on both the heavy pseudo-scalar mass and the chosen light scalar benchmark. The significance remains sizable in the low to intermediate mass region, reaching values well above the $5\sigma$ level for favourable configurations; in particular, for $M_{h} = 50~\text{GeV}$, a mass of $M_{A} \sim 375~\text{GeV}$ can achieve a $5\sigma$ significance. Although the overall reach is reduced compared to the model-dependent case shown in \autoref{fig:ReachPlot}(b), the degradation remains moderate. For instance, for $M_{h} = 70~\text{GeV}$, the $2\sigma$ exclusion reach decreases from $M_{A} \sim 560~\text{GeV}$ with the $\mathcal{C}_2$ cut to about $M_{A} \sim 515~\text{GeV}$ in this setup. Despite the absence of explicit resonance-based selection, the $1\ell + 2J_{bb}$ channel retains a substantial discovery reach. This demonstrates that the essential features of the signal can be effectively captured through kinematic observables, highlighting the robustness of the model-independent strategy.

The right panel of \autoref{fig:RchPlt_MdlInd} shows the parameter space in the $M_A-\tb$ plane that can be explored in the model-independent analysis. The solid lines show 5$\sigma$ reach at HL-LHC for various light scalar masses, and the dot-dashed curves depict the exclusion limit at $2 \sigma$ for 300\,\fbi. The blue, green, and red curves correspond to $M_{h} = 30$, $50$, and $70~\rm{GeV}$, respectively. The horizontal shaded regions are for existing theoretical and collider bounds for three light scalar benchmark points mentioned in \autoref{sec:3-bp}. As before our signal can exclude the large $\tb$ region. However, the reach in $M_A$ is somewhat reduced compared to the model-dependent analysis (see \autoref{fig:tanb-mA}), reflecting the absence of the fat-jet mass window cut. Despite this, the model-independent approach retains significant sensitivity across a wide region of the parameter space.

\subsection{Reconstruction of Heavy Scalars}
Apart from the sensitivity of the signal in the model-independent framework, it is crucial to investigate whether the underlying structure of the signal can be identified without explicit mass-based selection. For this, we focus on the \emph{fully visible} $2\ell + 2J_{bb}$ final state and study whether the fat-jets containing two $b$-subjets from the light scalar decay retain their characteristic features.

\begin{figure}[t]
	\begin{center}
		\hspace*{-1.0cm}
		\mbox{\subfigure[]{\includegraphics[width=0.36\linewidth,angle=0]{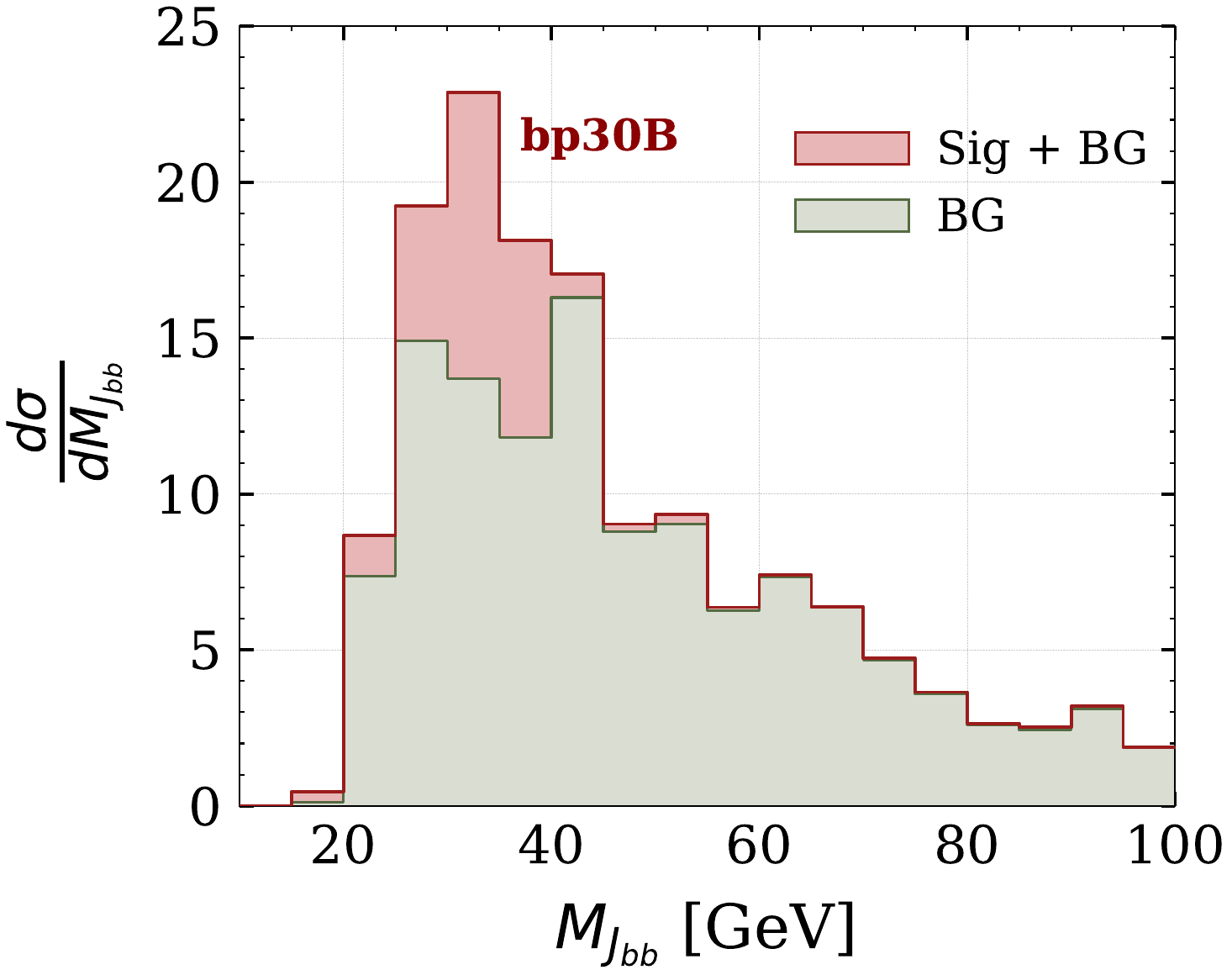}}
			\subfigure[]{\includegraphics[width=0.36\linewidth,angle=0]{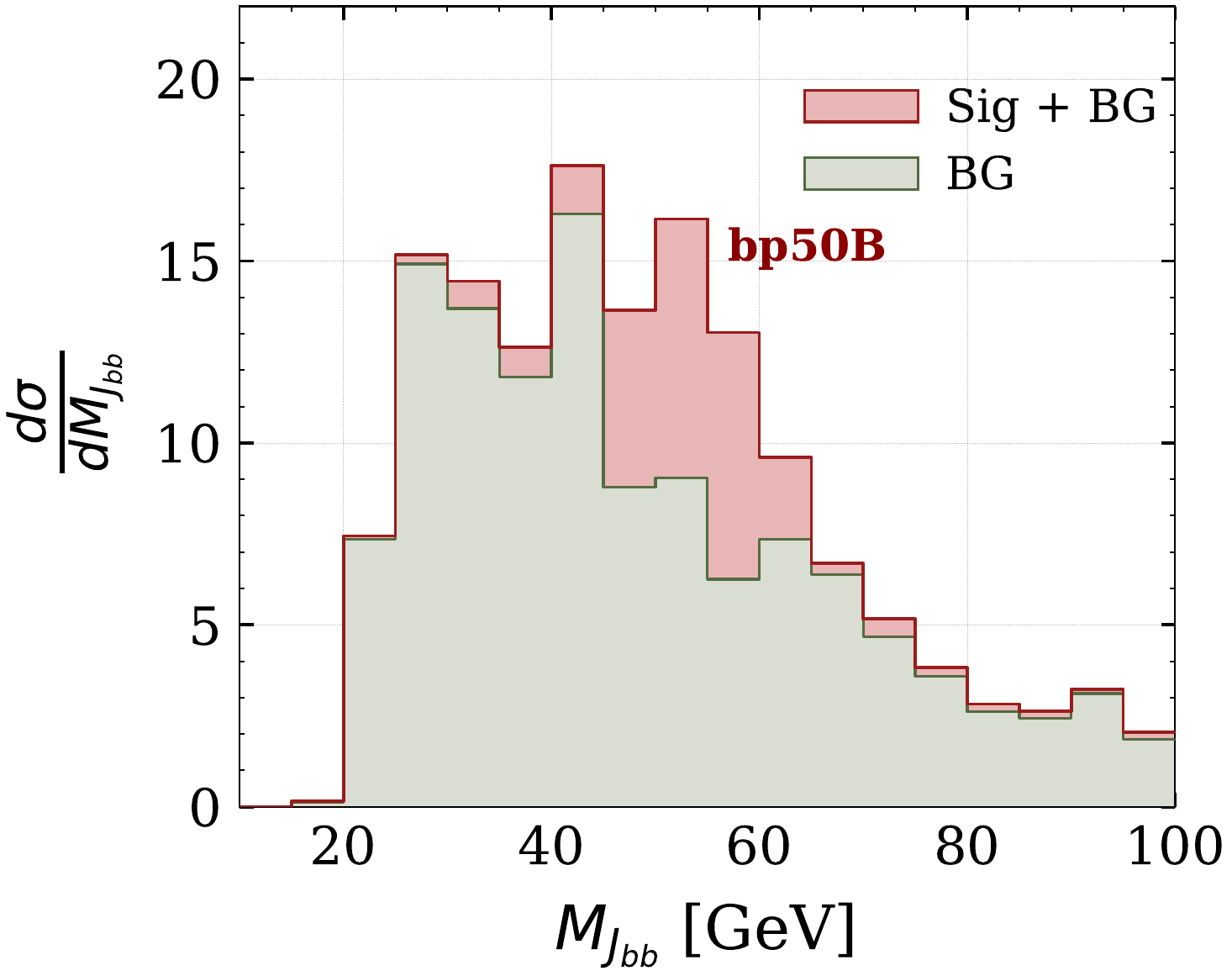}}
			\subfigure[]{\includegraphics[width=0.36\linewidth,angle=0]{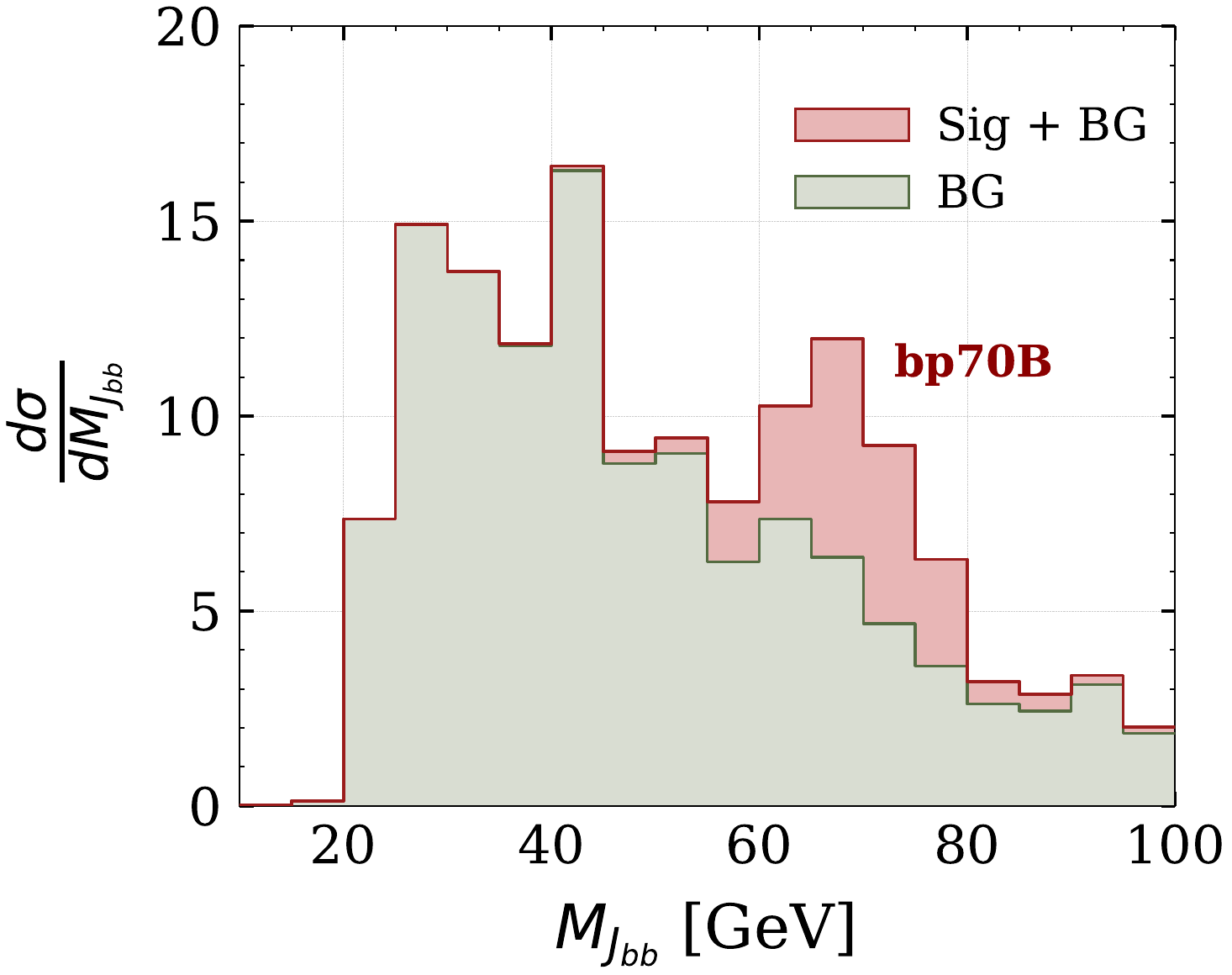}}}
	\caption{Distribution of the reconstructed fat-jet mass $M_{J_{bb}}$ for the $2\ell + 2J_{bb}$ final state in the model-independent analysis at the LHC with $\sqrt{s}=14~\rm{TeV}$ and an integrated luminosity of 3000\,\fbi. The signal (Sig + BG) and background (BG) contributions are shown for representative benchmark points (a) bp30B, (b) bp50B, and (c) bp70B.} \label{fig:FjMass}
	\end{center}
\end{figure}

The distribution of the fat-jet mass $M_{J_{bb}}$ for the highest-$p_T$ fat-jet is shown in \autoref{fig:FjMass} for representative benchmark points. These distributions are obtained after applying the cuts $\mathcal{C}_0$, $\mathcal{C}_1$, and $\mathcal{C}_3$ (see \autoref{Tab:FS4ind_Sig}), all of which are independent of the assumed light scalar mass. The SM background is shown in olive green, while the total contribution (signal + background) is displayed in red. A clear enhancement of the signal over the background is observed in the region corresponding to the light scalar mass for each benchmark point. In particular, distinct hump-like structures appear in the fat-jet mass distribution, which are absent in the background and arise from the boosted $h \to b\bar{b}$ decay within the fat-jet. This observation indicates that the signal exhibits a distinguishable structure from the background.


\begin{figure}[t]
	\begin{center}
		\hspace*{-1.0cm}
		\mbox{\subfigure[]{\includegraphics[width=0.36\linewidth,angle=0]{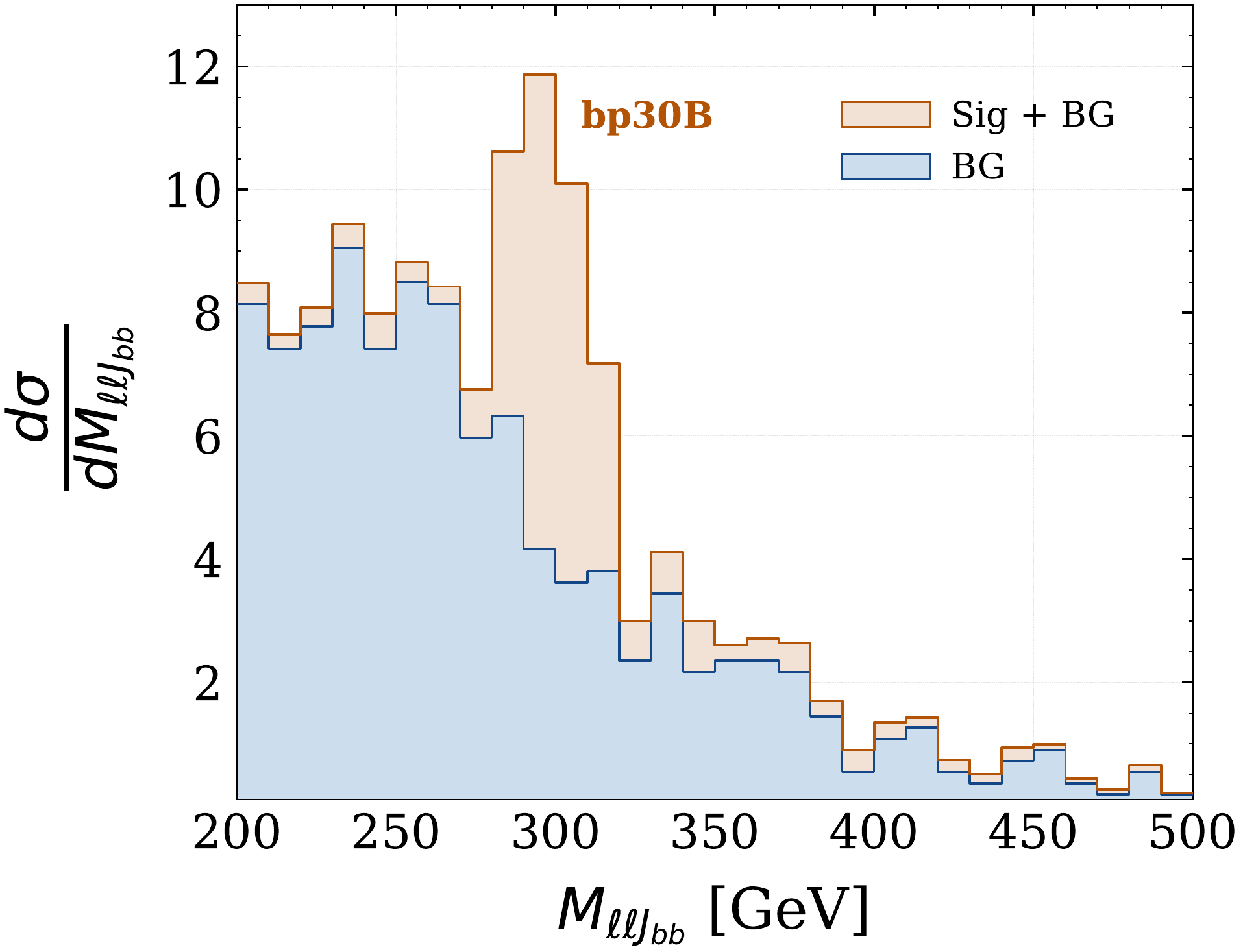}}
			\subfigure[]{\includegraphics[width=0.36\linewidth,angle=0]{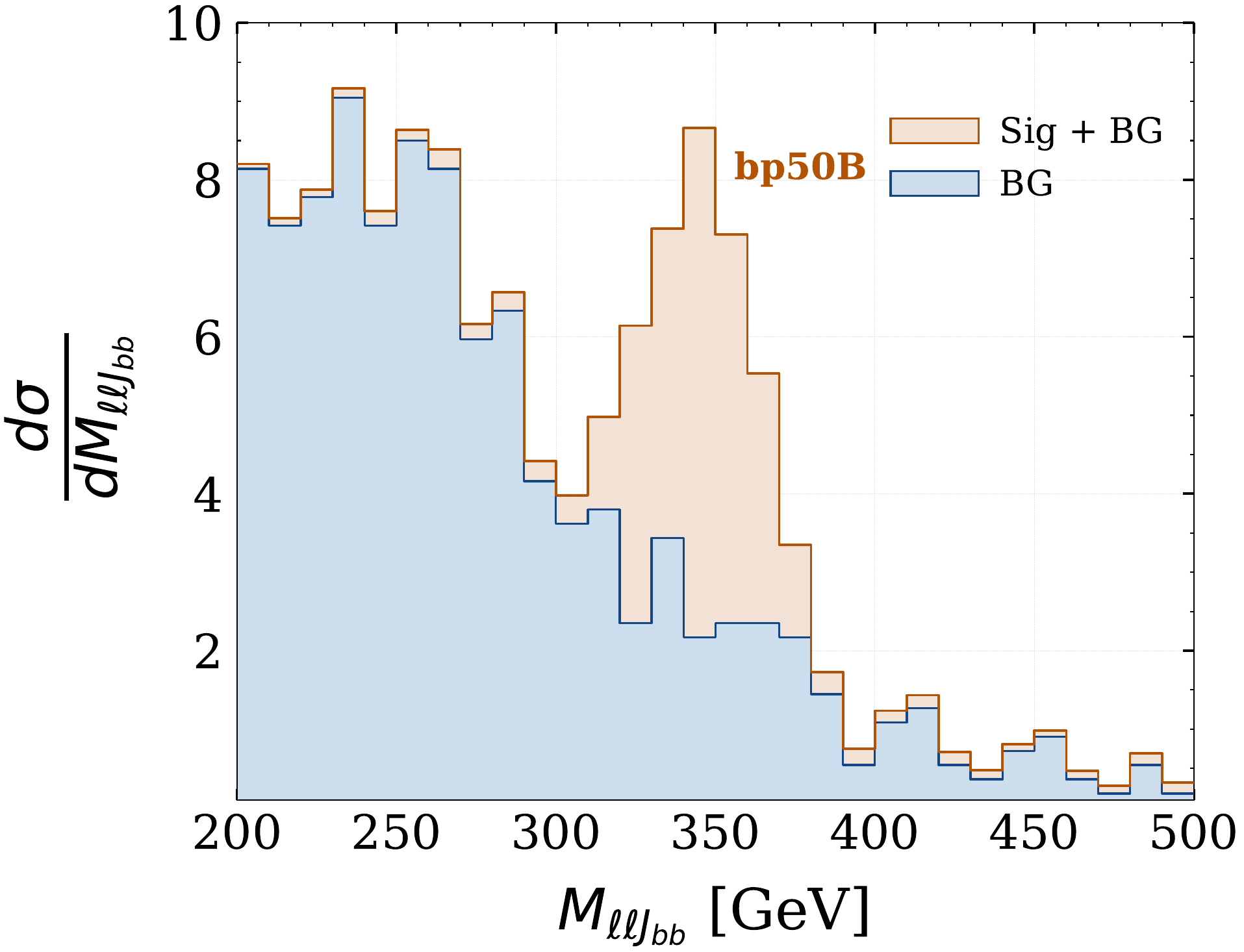}}
			\subfigure[]{\includegraphics[width=0.36\linewidth,angle=0]{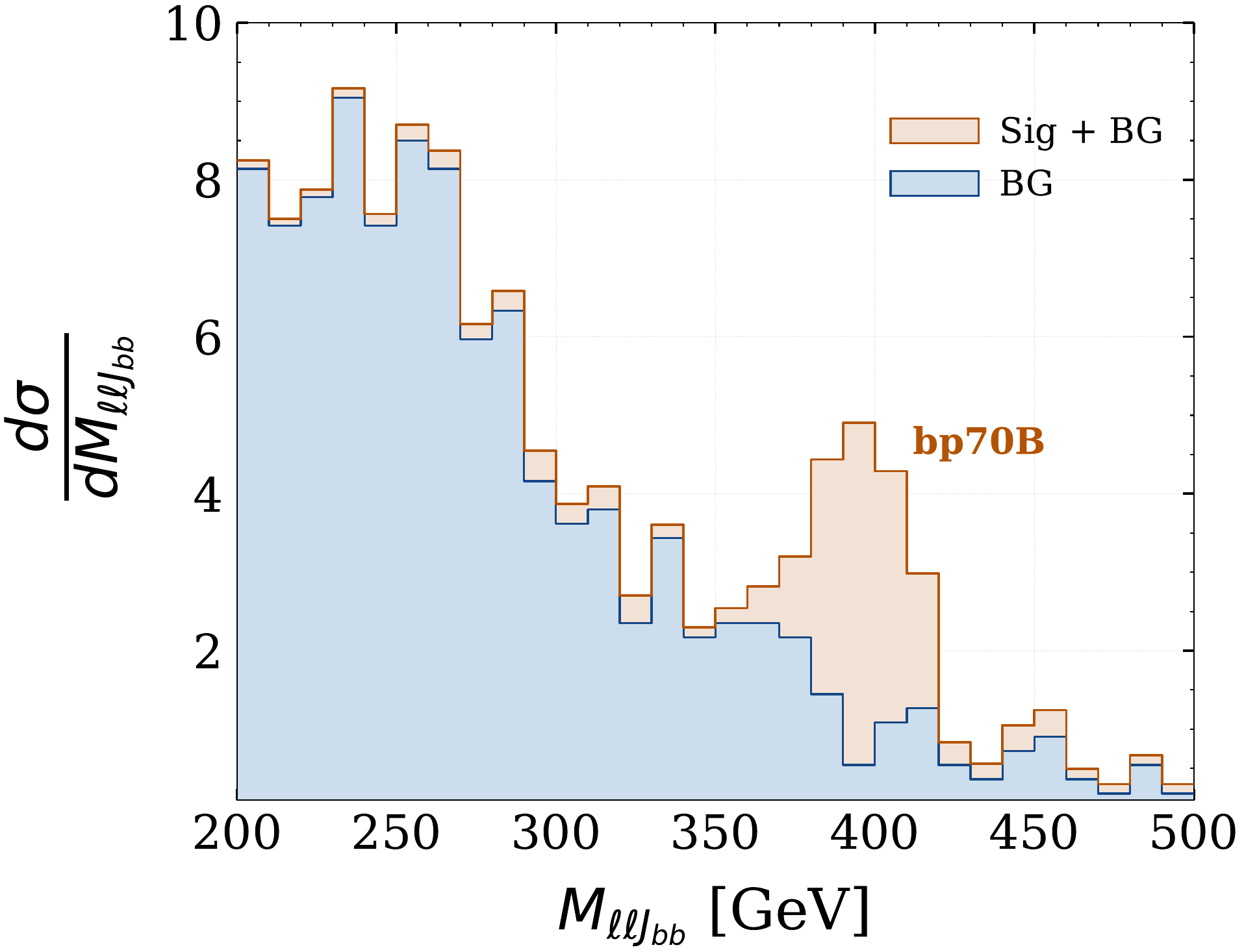}}}
		\caption{Distribution of the reconstructed invariant mass $M_{\ell\ell J_{bb}}$ for the $2\ell + 2J_{bb}$ final state in the model-independent analysis at the LHC with $\sqrt{s}=14~\rm{TeV}$ and an integrated luminosity of 3000~\fbi. The distributions are obtained after applying the cuts $\mathcal{C}_0$, $\mathcal{C}_1$, and $\mathcal{C}_3$, without imposing any explicit requirement on the fat-jet mass. The SM background is shown in light blue, while the total contribution (signal + background) is shown in orange. Panels (a)--(c) correspond to the benchmark points bp30B, bp50B, and bp70B, respectively, where the peaks in the $M_{\ell\ell J_{bb}}$ distribution align with the corresponding $A$ masses.} \label{fig:InvM}
	\end{center}
\end{figure}
We also explored whether the heavy scalar resonance can be reconstructed using only model-independent observables. 
The invariant mass $M_{\ell\ell J_{bb}}$ is constructed from the dilepton system and the highest-$p_T$ fat-jet, without imposing any explicit requirement on the fat-jet mass. The resulting distributions for representative benchmark points (bp30B, bp50B, and bp70B) are shown in \autoref{fig:InvM}. The light blue distribution represents the SM background, while the orange histogram corresponds to the total contribution including both signal and background. A clear excess over the background is observed, with peaks in the $M_{\ell\ell J_{bb}}$ distribution corresponding to the reconstructed $A$ masses. This demonstrates that even without the application of the model-dependent cut ($\mathcal{C}_2$), the boosted topology and event kinematics are sufficient to recover the underlying resonance structure.

\section{Summary and Conclusion}\label{sec:6-concl}

In this work, we investigate boosted double-$b$ fat-jet signatures as a probe of light scalars in a hierarchical mass spectrum within the Type-I 2HDM. The signal arises from electroweak production of scalar pairs, including the associated production modes $pp\to hA$ and $pp\to hH^\pm$, as well as the heavy-scalar pair-production modes $pp\to AH^\pm$ and $pp\to H^+H^-$. The subsequent decays $A\to Zh$ and $H^\pm\to W^\pm h$ lead to final states containing multiple light scalars together with electroweak gauge bosons. Owing to the large mass hierarchy, the light scalars are typically produced with large transverse momentum, leading to collimated decay products. Such a boosted regime emerges naturally from hierarchical scalar spectra. Hence, conventional resolved $b$-jet analyses cannot efficiently probe this region of the parameter space.

We have shown that requiring such large-radius jets, each containing two $b$-subjets, leads to a significant suppression of SM backgrounds. This allows us to explore heavy scalar masses beyond the reach of existing search strategies, extending well above $550~\text{GeV}$ at the high luminosity LHC.
We have performed a comprehensive analysis of several final states based on the multiplicity of leptons and boosted double-$b$ tagged fat-jets ($J_{bb}$), and found that the $1\ell + 2J_{bb}$ channel provides the best sensitivity. In addition to benchmark-based analyses, we have also explored a model-independent strategy that does not rely on the masses of BSM scalars. Although the benchmark analysis yields slightly higher significance, the model-independent approach remains robust, with sensitivity that is not far from the benchmark case. We have also demonstrated that it is indeed possible to reconstruct the resonance structure using kinematic observables even in the absence of mass-based selection criteria.

We have also mapped the LHC reach of the proposed analysis in the $M_A$--$\tan\beta$ plane. A large region at moderate to high $\tan\beta$ remains unconstrained by current experimental bounds but can be efficiently probed using our proposed $J_{bb}$ signature. Even at an integrated luminosity of 300~\fbi, a substantial part of this parameter space can be excluded at $2\sigma$, and the reach expands significantly at the high-luminosity LHC. Crucially, the reach of the proposed signal is largely independent of $\tan\beta$, allowing us to probe a wide region of the parameter space that is not accessible with existing search strategies.


\section*{Acknowledgements}
Research work at the Physical Research Laboratory (PRL) is funded by the Department of Space, Government of India. The authors gratefully acknowledge the resources provided by the Param Vikram-1000 High Performance Computing Cluster infrastructure at PRL, which were extensively used for the computational aspects of this work. T.M. was supported by BITS Pilani Seed Grant (NFSG/PIL/2023/P3801) and ANRF (ANRF/ECRG/2025/002387/PMS).

\appendix
\section{Cut Flow Tables for Additional Final States} \label{app:cutflow}
For completeness, we present the detailed cut-flow tables and corresponding discussion for the subleading final states considered in this work. These channels exhibit reduced sensitivity compared to the $1\ell + 2J_{bb}$ final state discussed in the main text.

\begin{table*}[hbt]
	\renewcommand{\arraystretch}{1.2}
	\centering
	\scalebox{0.95}
	{\begin{tabular}{|c||c|c|c|c|}
			\hline
			\multicolumn{5}{|c|}{Final State: $1 \ell + 1 J_{bb}$ }  \\ 
			\hline
			\multirow{2}{*}{Benchmark Points} & \multicolumn{4}{c|}{Cut flow}  \\ 
			\cline{2-5}
			& Modes  & $\mathcal{C}_0$  & $\mathcal{C}_2 $ &  Significance ($\sigma$)  \\ 
			\hline
			\multirow{3}{*}{bp30A}  & $h A$ & 6248.5 & 4591.7 &  \multirow{5}{*}{12.70}   \\ 
			& $h H^\pm$ & $3.07 \times 10^4$ & $1.71 \times 10^4$ &    \\
			& $A H^\pm$ & 9340.9 & 3516.5 &    \\
			& $H^+ H^-$ & 7072.3 & 2375.6 &    \\
			& BG & $2.47 \times 10^7$ & $4.71 \times 10^6$  &    \\
			\hline
			\multirow{3}{*}{bp30B}  & $h A$ & 963.2 & 678.4 &  \multirow{5}{*}{2.39}  \\ 
			& $h H^\pm$ & 6056.9 & 3716.9 &    \\
			& $A H^\pm$ & 1267.2 & 483.5 &    \\
			& $H^+ H^-$ & 968.8 & 304.0 &    \\
			& BG & $2.47 \times 10^7$ & $4.71 \times 10^6$ &    \\
			\hline
			\multirow{3}{*}{bp50A}  & $h A$ & 1885.1 & 1292.8 & \multirow{5}{*}{5.23}  \\
			& $h H^\pm$ & $1.07 \times 10^4$ & 7804.7 &    \\
			& $A H^\pm$ & 2823.5 & 1312.3 &    \\
			& $H^+ H^-$ & 2173.1 & 1034.6 &    \\
			& BG & $2.47 \times 10^7$ & $4.78 \times 10^6$  &    \\
			\hline
			\multirow{3}{*}{bp50B}  & $h A$ & 711.4 & 451.2 &   \multirow{5}{*}{1.91}  \\ 
			& $h H^\pm$ & 4453.6 & 2999.8 &    \\
			& $A H^\pm$ & 906.0 & 403.4 &    \\
			& $H^+ H^-$ & 725.0 & 321.7 &    \\
			& BG & $2.47 \times 10^7$ & $4.78 \times 10^6$  &   \\
			\hline
			\multirow{3}{*}{bp70A}  & $h A$ & 947.9 & 559.6 &  \multirow{5}{*}{4.99}   \\
			& $h H^\pm$ & 5529.5 & 3514.0 &   \\
			& $A H^\pm$ & 1450.1 & 645.9 &    \\
			& $H^+ H^-$ & 1111.7 & 544.6 &    \\
			& BG & $2.47 \times 10^7$ & $1.11 \times 10^6$  &    \\
			\hline
			\multirow{3}{*}{bp70B}  & $h A$ & 424.8 & 249.5 &  \multirow{5}{*}{2.30}  \\ 
			& $h H^\pm$ & 2703.3 & 1689.7 &    \\
			& $A H^\pm$ & 557.8 & 255.8 &    \\
			& $H^+ H^-$ & 442.1 & 224.3 &    \\
			& BG & $2.47 \times 10^7$  &  $1.11 \times 10^6$  &   \\
			\hline
	\end{tabular}}
	\caption{Event yields and signal significance for the $1\ell + 1J_{bb}$ final 	state at the LHC with $\sqrt{s}=14~\rm{TeV}$ and an integrated luminosity of  3000 \fbi. The contributions from the $hA$, $hH^\pm$, $AH^\pm$ and $H^+ H^-$ signal processes, as well as the total background (BG), are shown after the baseline selection ($\mathcal{C}_0$) and the fat-jet mass window cut ($\mathcal{C}_2$). The resulting signal significance ($\sigma$) is reported for each benchmark point.}  \label{Tab:FS1_sig}
\end{table*}
\subsection*{\textbf{\boldmath $1\ell + 1J_{bb}$}}

The cut-flow results for the $1\ell + 1J_{bb}$ final state at the LHC with the centre-of-mass energy of 14\,TeV and an integrated luminosity of 3000 \fbi are presented in \autoref{Tab:FS1_sig}. Starting from the baseline selection ($\mathcal{C}_0$), we observe that the background contribution is overwhelmingly large compared to the signal. The application of the fat-jet mass window cut ($\mathcal{C}_2$) leads to a substantial reduction in the background, while retaining a significant fraction of the signal events. This demonstrates the effectiveness of reconstructing the boosted scalar through the fat-jet mass as a key discriminating observable, particularly in final states with minimal particle multiplicity where the $\mathcal{C}_1$ and $\mathcal{C}_3$ cuts cannot be applied.

All four signal production modes, $hA$, $hH^\pm$, $AH^\pm$ and $H^+H^-$, contribute to the $1\ell+1J_{bb}$ event yield. Among these, the $hH^\pm$ production mode consistently gives the dominant contribution, owing to its larger production cross-section and the presence of a leptonically decaying $W^\pm$ boson in the final state. The channels containing a $Z$ boson in the decay chain, namely $hA$ and $AH^\pm$, can also populate the single-lepton category when one of the leptons from the $Z$ decay is not reconstructed or falls outside the detector acceptance. In addition, $AH^\pm$ and $H^+H^-$ can contribute through leptonic decays of the charged gauge boson. These heavy-scalar pair-production modes provide additional but comparatively smaller contributions to the total signal yield.

The resulting signal significance varies across the benchmark points, with the highest sensitivity achieved for lighter mass configurations, such as bp30A, where the boosted topology enhances the reconstruction efficiency. The degree of boost is governed by the mass difference between the heavy scalars ($A$/$H^\pm$) and the light scalar ($h$). For heavier benchmark points, although the boost effect becomes more pronounced, the reduction in production cross-section leads to an overall decrease in the signal significance. Overall, the $1\ell + 1J_{bb}$ channel provides moderate sensitivity, but remains limited by the large residual background, motivating the exploration of cleaner final states in the subsequent analysis.


\begin{table*}[t]
	\renewcommand{\arraystretch}{1.3}
	\centering
	\scalebox{0.95}
	{\begin{tabular}{|c||c|c|c|c|c|}
			\hline
			\multicolumn{6}{|c|}{Final State: $2 \ell + 1 J_{bb}$ }  \\ 
			\hline
			\multirow{2}{*}{Benchmark Points} & \multicolumn{5}{c|}{Cut flow}  \\ 
			\cline{2-6}
			& Modes  & $\mathcal{C}_0$  & $\mathcal{C}_1 $ & $\mathcal{C}_2 $  & Significance ($\sigma$)  \\ 
			\hline
			\multirow{2}{*}{bp30A}  & $h A$ & 3697.0 & 3255.6 & 2526.7 &  \multirow{4}{*}{8.86}  \\ 
			& $A H^\pm$ & 1512.6 & 1013.3 & 312.1 &  \\
			& $H^+ H^-$ & 641.0 & 71.5 & 43.6 &    \\
			& BG & $8.68 \times 10^5$ & $6.72 \times 10^5$ & $1.05 \times 10^5$ &    \\
			\hline
			\multirow{2}{*}{bp30B}  & $h A$ & 823.6 & 691.5 & 438.8 &  \multirow{4}{*}{1.55}  \\ 
			& $A H^\pm$ & 254.4 & 172.1 & 58.4 &  \\
			& $H^+ H^-$ & 120.0 & 9.0 & 5.9 &    \\
			& BG & $8.68 \times 10^5$  & $6.72 \times 10^5$  &  $1.05 \times 10^5$ &    \\
			\hline
			\multirow{2}{*}{bp50A}  & $h A$ &  1350.1 & 1122.4  & 719.8  &   \multirow{4}{*}{2.74}  \\
			& $A H^\pm$ & 545.3 & 353.9 & 129.8 &  \\
			& $H^+ H^-$ & 273.2 & 26.3 & 17.3 &    \\
			& BG & $8.68 \times 10^5$  & $6.72 \times 10^5$  & $9.99 \times 10^4$  &    \\
			\hline
			\multirow{2}{*}{bp50B}   & $h A$ & 591.1 & 473.9 & 250.4 &   \multirow{4}{*}{0.93}  \\ 
			& $A H^\pm$ & 196.6 & 126.4 & 40.1 &  \\
			& $H^+ H^-$ & 98.1 & 6.6 & 4.2 &    \\
			& BG & $8.68 \times 10^5$  & $6.72 \times 10^5$  & $9.99 \times 10^4$ &   \\
			\hline
			\multirow{2}{*}{bp70A}  & $h A$ & 676.1 & 551.9 & 292.5  &  \multirow{4}{*}{2.60}   \\
			& $A H^\pm$ & 288.3 & 184.5 & 68.5 &  \\
			& $H^+ H^-$ & 142.8 & 12.0 & 7.4 &    \\
			& BG &  $8.68 \times 10^5$ & $6.72 \times 10^5$  & 20026.4 &    \\
			\hline
			\multirow{2}{*}{bp70B}   & $h A$ & 352.1 & 281.0 & 131.6 & \multirow{4}{*}{1.14}  \\ 
			& $A H^\pm$ & 122.7 & 76.0 & 27.3 &  \\
			& $H^+ H^-$ & 61.6 & 3.9 & 2.5 &    \\
			& BG &  $8.68 \times 10^5$ & $6.72 \times 10^5$  & 20026.4 &   \\
			\hline
	\end{tabular}}
	\caption{Event yields and signal significance for the $2\ell + 1J_{bb}$ final 
	state at $\sqrt{s}=14~\text{TeV}$ and an integrated luminosity of 3000~\fbi. The contributions from the $hA$, $AH^\pm$ and $H^+H^-$ signal processes, as well as the total background (BG) are shown after the baseline selection ($\mathcal{C}_0$), the dilepton invariant mass cut ($\mathcal{C}_1$), and the fat-jet mass window cut ($\mathcal{C}_2$).}  \label{Tab:FS3_sig}
\end{table*}
\subsection*{\textbf{\boldmath $2\ell + 1J_{bb}$}}

The cut flow for the $2\ell + 1J_{bb}$ final state is shown in \autoref{Tab:FS3_sig}. In contrast to the single-lepton channels, this final state is primarily sensitive to production modes containing a $Z$ boson in the decay chain. The dominant contribution therefore comes from the $hA$ production mode, where the $Z$ boson decays leptonically. The heavy-scalar pair-production mode $AH^\pm$ also contributes, since the decay $A\to hZ$ can give rise to the same dilepton signature. The $H^+H^-$ contribution is comparatively smaller: although it can produce two leptons through leptonic $W^+W^-$ decays, these leptons do not originate from an on-shell $Z$ boson and are therefore strongly reduced by the dilepton invariant-mass requirement $\mathcal{C}_1$.

The application of the dilepton invariant mass cut $\mathcal{C}_1$ effectively suppresses backgrounds by selecting events consistent with on-shell $Z$ boson decays. This, combined with the fat-jet mass window cut $\mathcal{C}_2$, provides a cleaner environment compared to the single-lepton case. However, the absence of a second fat-jet prevents the use of the mass-asymmetry cut $\mathcal{C}_3$, which was a key discriminator in the $1\ell+2J_{bb}$ channel. As a result, residual background contributions remain non-negligible.

The resulting signal significance in this channel is therefore lower than that of the $1\ell + 2J_{bb}$ final state. While the cleaner dilepton signature helps in reducing background contamination, the reduced signal yield and the lack of additional topological discrimination limit the achievable sensitivity. This channel nonetheless provides a complementary probe of the model, particularly for validating the $hA$ and $AH^\pm$ production modes in a cleaner experimental environment.


\begin{table*}[hbt]
	\renewcommand{\arraystretch}{1.3}
	\centering
	\scalebox{0.95}
	{\begin{tabular}{|c||c|c|c|c|c|c|}
			\hline
			\multicolumn{7}{|c|}{Final State: $2 \ell + 2 J_{bb}$ }  \\ 
			\hline
			\multirow{2}{*}{Benchmark Points} & \multicolumn{6}{c|}{Cut flow}  \\ 
			\cline{2-7}
			& Modes  & $\mathcal{C}_0$  & $\mathcal{C}_1 $ & $\mathcal{C}_2 $ & $\mathcal{C}_3 $ & Significance ($\sigma$)  \\ 
			\hline
			\multirow{2}{*}{bp30A}  & $h A$ & 188.8 & 175.6 & 158.1 & 36.7  & \multirow{4}{*}{3.43}  \\ 
			& $A H^\pm$ & 162.4 & 109.9  & 48.5  & 9.9 &  \\
			& $H^+ H^-$ & 55.3  & 7.1  & 6.0  & 2.5 &  \\
			& BG & 3397.6 & 2821.7 & 1575.6 & 188.8 &   \\
			\hline
			\multirow{2}{*}{bp30B}  & $h A$ & 114.8  & 101.8 & 96.8 & 44.8 & \multirow{4}{*}{3.45}  \\ 
			& $A H^\pm$ & 29.1  & 21.5 & 15.0 & 4.2 &  \\
			& $H^+ H^-$ & 11.9  & 0.9  & 0.8 & 0.4 &  \\
			& BG & 3397.6 & 2821.7 & 1575.6 & 188.8 &   \\
			\hline
			\multirow{2}{*}{bp50A}  & $h A$ & 222.1  & 203.1 & 190.9 & 108.8  &  \multirow{4}{*}{6.07}  \\
			& $A H^\pm$ & 87.0 & 62.0  & 46.1 & 20.2 &  \\
			& $H^+ H^-$ & 42.0 & 4.9 & 4.6 & 2.6 &  \\
			& BG & 3397.6 & 2821.7 & 2289.9 & 428.2  &   \\
			\hline
			\multirow{2}{*}{bp50B}   & $h A$ & 136.9 & 121.4 & 111.8 & 55.9  & \multirow{4}{*}{3.04}  \\ 
			& $A H^\pm$ & 42.6  & 31.0 & 22.7 & 7.7 &  \\
			& $H^+ H^-$ & 18.6  & 1.4 & 1.3 & 0.8 &  \\
			& BG & 3397.6 & 2821.7 & 2289.9 & 428.2 &   \\
			\hline
			\multirow{2}{*}{bp70A}  & $h A$ & 100.7 & 90.5 & 79.8 & 50.3 & \multirow{4}{*}{4.29}   \\
			& $A H^\pm$ &  43.2 & 30.2 & 21.9  & 12.1 &  \\
			& $H^+ H^-$ & 20.6 & 1.8 & 1.7  & 1.2 &  \\
			& BG & 3397.6 & 2821.7 & 1516.4 & 199.4 &   \\
			\hline
			\multirow{2}{*}{bp70B} & $h A$ & 74.1 & 65.9 & 57.6 & 36.1 & \multirow{4}{*}{3.00}  \\ 
			& $A H^\pm$ & 24.5  & 17.2  & 13.0  & 7.3 &  \\
			& $H^+ H^-$ & 11.7  & 0.8  & 0.7  & 0.5 &  \\
			& BG & 3397.6 & 2821.7 & 1516.4 & 199.4 &   \\
			\hline
	\end{tabular}}
	\caption{Event yields and signal significance for the $2\ell + 2J_{bb}$ final state at the LHC with $\sqrt{s}=14~\rm{TeV}$ and an integrated luminosity of 3000~\fbi, obtained using the same selection strategy as in \autoref{Tab:FS3_sig}, with the inclusion of the fat-jet mass asymmetry cut ($\mathcal{C}_3$).}  \label{Tab:FS4_sig}
\end{table*}
\subsection*{\textbf{\boldmath $2\ell + 2J_{bb}$}}
We finally consider the $2\ell + 2J_{bb}$ final state at the LHC with the centre-of-mass energy of 14 TeV and an integrated luminosity of 3000~\fbi, and the corresponding results are shown in \autoref{Tab:FS4_sig}. This channel represents the cleanest topology among all the signal regions, as it combines the dilepton requirement with the reconstruction of two boosted fat-jets. Consequently, the background is significantly reduced compared to all previously considered channels. 
The successive application of the cuts, particularly the dilepton invariant mass cut ($\mathcal{C}_1$) and the fat-jet mass asymmetry cut ($\mathcal{C}_3$), leads to a strong suppression of the background, resulting in the smallest background yields among all final states. However, this stringent event selection also significantly reduces the signal statistics. In this channel, the dominant contribution arises from the $hA$ mode, while $AH^\pm$ provides an additional subleading contribution through the $A\to hZ$ decay. The $H^+H^-$ contribution is comparatively small, since the dileptons originate from leptonic $W^+W^-$ decays and are strongly reduced by the $Z$-mass requirement $\mathcal{C}_1$.

As a result, despite the very clean final state, the signal significance does not exceed that of the $1\ell + 2J_{bb}$ channel. This highlights the trade-off between background suppression and signal statistics: while tighter selection criteria lead to a cleaner environment, they can also reduce the signal yield to a level where the overall sensitivity is diminished. This channel remains important as a robust and clean confirmation mode, providing a fully visible and reconstructible final state for the signal.

\bibliographystyle{JHEP}
\bibliography{reference}
\end{document}